\documentclass[%
  reprint,
  onecolumn,
  amsmath,
  amssymb,
  10pt,
  aps,
  prx,
  citeautoscript,
  notitlepage,
  longbibliography
]{revtex4-2}

\usepackage[utf8]{inputenc}
\usepackage[T1]{fontenc}
\usepackage{xcolor}
\definecolor{blue}{RGB}{50,50,220}
\usepackage[%
  colorlinks=true,
  allcolors=blue
]{hyperref}
\usepackage{url}
\usepackage{enumitem}
\usepackage{amsfonts}
\usepackage{amssymb}
\usepackage{amsmath}
\usepackage{mathtools}
\usepackage[ruled,vlined]{algorithm2e}
\usepackage{lmodern}
\usepackage{parskip}
\usepackage[left=3cm,right=3cm]{geometry}

\DeclarePairedDelimiter\ppar{(}{)}              
\DeclarePairedDelimiter\pang{\langle}{\rangle}  
\DeclarePairedDelimiter\pnrm{\lVert}{\rVert}    
\DeclarePairedDelimiter\pbkt{[}{]}              
\DeclarePairedDelimiter\pset{\{}{\}}            

\newcommand{\rfig}[1]{Fig.~\ref{#1}}

\newcommand{\rsct}[1]{Sec.~\ref{#1}}
\newcommand{\rref}[1]{Ref.~\citenum{#1}}
\newcommand{\req}[1]{Eq.~\ref{#1}}

\newcommand{\dd}[1]{\operatorname{d#1}}

\newcommand{\bz}{\mathbf{z}}
\newcommand{\bx}{\mathbf{x}}
\newcommand{\by}{\mathbf{y}}

\newcommand{\dz}{\dd{\mathbf{z}}}
\newcommand{\dx}{\dd{\mathbf{x}}}
\newcommand{\dy}{\dd{\mathbf{y}}}
\newcommand{\btheta}{\pmb{\theta}}
\newcommand{\e}{\operatorname{e}}
\newcommand{\kT}{k_{\mathrm{B}}T}

\newcommand{\cM}{\mathcal{M}}
\newcommand{\cL}{\mathcal{L}}
\newcommand{\cK}{\mathcal{K}}

\newcommand{\chg}[1]{{\color{black}{#1}}}
\usepackage[normalem]{ulem}

\begin{document}
\title{%
  Manifold Learning in Atomistic Simulations:\\
  A Conceptual Review
}

\author{Jakub Rydzewski}
\email[]{jr@fizyka.umk.pl}
\affiliation{%
  Institute of Physics,
  Faculty of Physics, Astronomy and Informatics,
  Nicolaus Copernicus University,
  Grudziadzka 5, 87-100 Toru\'n, Poland
}

\author{Ming Chen}
\affiliation{%
  Department of Chemistry,
  Purdue University,
  West Lafayette, Indiana 47907-2048, USA
}

\author{Omar Valsson}
\affiliation{%
  Department of Chemistry,
  University of North Texas,
  Denton, Texas 76201, USA
}


\begin{abstract}
Analyzing large volumes of high-dimensional data requires dimensionality reduction: finding meaningful low-dimensional structures hidden in their high-dimensional observations. Such practice is needed in atomistic simulations of complex systems where even thousands of degrees of freedom are sampled. An abundance of such data makes gaining insight into a specific physical problem strenuous. Our primary aim in this review is to focus on unsupervised machine learning methods that can be used on simulation data to find a low-dimensional manifold providing a collective and informative characterization of the studied process. Such manifolds can be used for sampling long-timescale processes and free-energy estimation. We describe methods that can work on datasets from standard and enhanced sampling atomistic simulations. Unlike recent reviews on manifold learning for atomistic simulations, we consider only methods that construct low-dimensional manifolds based on Markov transition probabilities between high-dimensional samples. We discuss these techniques from a conceptual point of view, including their underlying theoretical frameworks and possible limitations.
\end{abstract}

\maketitle

\newpage
\tableofcontents
\newpage

\section{Introduction}
\label{sec:introduction}
Atomistic simulations are extensively used to investigate complex systems in chemistry and biology~\cite{10.1146/annurev-biophys-042910-155245,10.1126/science.aaz3041}. These simulations provide detailed information about physical and chemical processes at the atomistic level of detail with spatiotemporal resolution inaccessible to experiments. However, such systems often involve hundreds of thousands of atoms, making it challenging to analyze their high-dimensional configuration space. It requires techniques for averaging over noisy variables that correspond to fast degrees of freedom while obtaining a low-dimensional description that retains the essential characteristics of the associated physical processes. A low-dimensional representation should be physically explainable and interpretable. Paraphrasing R. Coifman~\cite{coiman2018icm}:
\begin{quote}
  There is innate truth in the low-dimensional manifold of the data, and we would like to have a characterization of some latent variable that intrinsically describes the changes of states,
\end{quote}
we can intuitively understand the motivation to develop methods for finding low-dimensional representations in atomistic simulations.

To better understand complex systems, they are studied using frameworks that can alleviate the apparent problem of high dimensionality~\cite{chandler1987introduction,coifman2005geometric,mezic2005spectral,valsson2016enhancing,wu2017variational,klus2018data,glielmo2021unsupervised,lin2021data,morishita2021time}. These frameworks include various approaches such as the Ginzburg--Landau theory of phase transitions~\cite{hohenberg2015introduction}, the Mori--Zwanzig formalism for transport and collective motion~\cite{zwanzig1961memory,luttinger1964theory,mori1965transport}, and Koopman's theory~\cite{wu2020variational,brunton2021modern}. More recently developed approaches include manifold learning~\cite{borg2005modern,lee2007nonlinear,van2009dimensionality,abdi2010principal,ma2012manifold,izenman2012introduction}, a class of nonlinear unsupervised machine learning methods trained directly on collected data, whose development was instigated by the innovative works of Tenenbaum et al.~\cite{tenenbaum2000global} and Roweis and Saul~\cite{roweis2000nonlinear}; both published in the same issue of Science [\textbf{290} (2000)].

Complex systems require a strict approach that ensures their informative physical characteristics are encoded in corresponding low-dimensional manifolds. In the context of atomistic simulations, encoding essential characteristics of the physical process while averaging over remaining degrees of freedom should be performed according to several requirements~\cite{valsson2016enhancing,noe2017collective,pietrucci_strategies_2017,bussi2007accurate}:
\begin{enumerate}[leftmargin=0.5cm]
  \item Distinguishing between relevant states of the system.
  \item Including slowly varying degrees of freedom corresponding to system behavior on longer timescales.
\end{enumerate}%

Another difficulty in finding low-dimensional manifolds from atomistic simulations arises due to the sampling problem. Complex systems are often characterized by metastable states separated by energy barriers much higher than thermal energy $\kT$. This metastability leads to kinetic entrapment in a single state, making transitions between metastable states infrequent (i.e., rare). As a result, metastable systems are sampled only in a fraction of their configuration space due to the low probability of jumping across energy barriers, and the data obtained cannot represent the whole behavior of the system. To address this issue, enhanced sampling methods can be used to bias the equilibrium probability and improve the sampling of the configuration space~\cite{valsson2016enhancing,bussi2020using,henin2022enhanced}. However, this problem is often overlooked, and many methods remain unable to learn from data generated by biased sampling.

\chg{The main objective of this review is to establish a theoretical framework for manifold learning methods suitable for investigating systems through atomistic simulations, including enhanced sampling simulations. However, we deviate from a commonly taken route to reviewing unsupervised learning methods for finding low-dimensional representations of complex systems. We do not discuss standard techniques such as principal component analysis, multidimensional scaling, and their variants, as these have already been covered in many reviews. For an introduction to the methods omitted here, we refer to reviews focusing on learning from machine learning datasets~\cite{borg2005modern,lee2007nonlinear,van2009dimensionality,ma2012manifold,izenman2012introduction,xie2020representation} or simulation data~\cite{noe2017collective,sittel2018perspective,ceriotti2019unsupervised,wang2020machine,bernetti2020data,noe2020machine,gkeka2020machine,glielmo2021unsupervised,chen2021collective,bhatia2023confluence}. 

Instead, we focus solely on a group of nonlinear techniques that construct Markov transition probabilities between high-dimensional samples. Recent development has shown that these techniques can be considered in one general framework suitable for complex systems sampled using atomistic simulations. \chg{For this reason, we consequently use the nomenclature employed in statistical physics and atomistic simulations.} Apart from using such manifold learning techniques for unbiased simulations, we also introduce concepts that allow the construction of low-dimensional manifolds from enhanced sampling simulations, in which the crucial information about the manifold is sampled from biased probability distributions. Overall, we review manifold learning methods under a single unifying framework, covering methods required to understand the latest developments in the field. In general, each algorithm we review here involves the following steps:
\begin{enumerate}[leftmargin=0.5cm]
  \item Generation of high-dimensional samples from unbiased or biased atomistic simulations.
  \item Construction of a Markov chain on the data with pairwise transition probabilities between samples.
  \item Parametrization of a manifold using a mapping that embeds high-dimensional samples to a reduced space through eigendecomposition (i.e., spectral embeddings~\cite{sha2005analysis}) or divergence optimization.
\end{enumerate}
}

This review begins with relatively standard material about atomistic simulations and a general introduction to enhanced sampling techniques (\rsct{sec:atomistic-simulations}). We cover only the concepts required to understand manifold learning in the context of standard atomistic and enhanced sampling simulations. This part, by no means exhaustive, can be supplemented by several comprehensive reviews on enhanced sampling~\cite{valsson2016enhancing,pietrucci_strategies_2017,yang2019enhanced,bussi2020using,kamenik2021enhanced,henin2022enhanced}. Next, a general problem of finding low-dimensional manifolds for the description of complex systems is introduced (\rsct{sec:manifold-learning}). Subsequently, we move to the central part of this review and focus on several manifold learning techniques that can be used for learning from atomistic simulations. Each of these frameworks is introduced from the conceptual perspective, followed by examples of applications and software implementations (\rsct{sec:tm-eigendecomposition} and \ref{sec:tm-divergence-optimization}). Finally, we summarize ongoing issues and provide our perspective on manifold learning in standard atomistic and enhanced sampling simulations (\rsct{sec:conclusions}).

\section{Atomistic Simulations}
\label{sec:atomistic-simulations}
Atomistic simulation techniques such as molecular dynamics or Monte Carlo have emerged as general methods at the intersection of theoretical and computational physics~\cite{battimelli2020computer}. These techniques used to explore the dynamics of complex systems can be viewed as \emph{samplers} for generating high-dimensional data from some underlying probability distributions.

\subsection{Statistical Physics Representation}
\label{sec:high-dimensional-space}
In statistical physics, we represent the dynamics of a complex system using its microscopic configurations. Such a representation generally involves a high number of degrees of freedom. Let us suppose that the system is represented by an $n$-dimensional vector of \emph{configuration variables}:
\begin{equation}
  \label{eq:configuration-variable}
  \bx \equiv \pset{x_k}_{k=1}^n = \ppar*{x_1, x_2, \dots, x_n},
\end{equation}
such as the \emph{microscopic coordinates}, where $n=3N$ for an $N$-atom system. Generally, the configuration variables are functions of the microscopic coordinates under the assumption that the space spanned by these variables is high-dimensional. In machine learning, the configurational variables are referred to as \emph{features} or \emph{descriptors}. A dataset of $K$ high-dimensional samples of the configuration variables:
\begin{equation}
  \label{eq:trajectory}
  X \equiv \pset{\bx_k}_{k=1}^K =
  \ppar*{\bx_1, \bx_2, \dots, \bx_K}
\end{equation}
recorded at consecutive times during the dynamics can be expressed as a matrix of size $n \times K$ called a \emph{trajectory}.


In the following, we limit our discussion to the canonical ensemble ($NVT$), in which the configuration variables evolve according to a high-dimensional potential energy function $U(\bx)$ at a temperature $T$. When the system is represented by the microscopic coordinates, its equilibrium density is given by the stationary Boltzmann distribution~\cite{chandler1987introduction}:
\begin{equation}
  \label{eq:boltzmann-density}
  \rho(\bx) = \frac{1}{\mathcal{Z}} \e^{-\beta U(\bx)},
\end{equation}
where $\beta=(\kT)^{-1}$ is the inverse of the thermal energy $\kT$ corresponding to the temperature $T$ with the Boltzmann constant denoted by $k_\mathrm{B}$, and $\mathcal{Z}=\int\dx\e^{-\beta U(\bx)}$ is the canonical partition function. Otherwise, the set of samples $X$ (\req{eq:trajectory}) is sampled from an unknown high-dimensional equilibrium density.


%
\begin{table*}
  \caption{Representations of a system considered in this review. Note that for simplicity, the same symbol $\bx$ is used for both the microscopic coordinates and configuration variables (features). See \rsct{sec:high-dimensional-space} for an explanation.}
  \begin{tabular}{p{3.8cm}p{0.6cm}p{4cm}p{2.2cm}p{2.5cm}c}
    \hline
    Variables &  & Probability distribution & Type & Dimensionality & \\
    \hline\hline
    Microscopic coordinates & $\bx$ & $\rho(\bx) \propto \e^{-\beta U(\bx)}$ & Equilibrium & High & $n$ \\
    Configuration variables & $\bx$ & $\rho(\bx)$ unknown & Equilibrium & High & $n$ \\
    Collective variables & $\bz$ & $\rho(\bz) \propto \e^{-\beta F(\bz)}$ & Equilibrium & Low & $d$ \\
    Collective variables & $\bz$ & $\rho_V(\bz) \propto \e^{-\beta (F(\bz)+V(\bz))}$ & Biased & Low & $d$ \\
    \hline
  \end{tabular}
\end{table*}

\subsection{Collective Variables and Target Mapping}
\label{sec:collective-variables}
The primary assumption in statistical physics is that we can average over some properties of the high-dimensional representation and obtain a macroscopic description of the system with fewer degrees of freedom that are capable of characterizing ensembles of the microscopic configurations known as states. In atomistic simulations, such macroscopic variables are often called \emph{collective variables} (CVs), order parameters, or reaction coordinates. Such variables can be considered as CVs for the system if they~\cite{abrams2014enhanced,valsson2016enhancing,pietrucci_strategies_2017,noe2017collective,sittel2018perspective,bussi2020using,neha2022collective}:
\begin{enumerate}[leftmargin=0.5cm]
  \item Encode information about essential characteristics.
  \item Distinguish between relevant states (i.e., modes).
  \item Include slowly varying degrees of freedom corresponding to long timescale processes.
\end{enumerate}
%

Identifying CVs is challenging for complex systems and often involves resorting to physical or chemical intuition and trial-and-error approaches~\cite{peters2016reaction}. This motivated many theoretical and computational advances to construct CVs directly from simulation data, for example, using neural networks~\cite{ma2005automatic,zhang2018unfolding,chen2018molecular,ribeiro2018reweighted,wehmeyer2018time,bonati2020data,sidky2020molecular,rydzewski2021multiscale,bonati2021deep,belkacemi2021chasing,rydzewski2022reweighted,ketkaew2022deepcv,jung2023machine}.

Let us assume for now that CVs are correctly identified through some procedure. CVs are typically expressed as functions of the configuration variables (\req{eq:configuration-variable}), meaning that finding CVs involves obtaining a set of functions that embed high-dimensional samples into a low-dimensional CV space. This set of functions is called the \emph{target mapping}~\cite{rydzewski2022reweighted}:
\begin{equation}
  \label{eq:collective-variables}
  \boxed{%
    \bx \mapsto \xi(\bx) \equiv \pset[\big]{ \xi_k(\bx) }_{k=1}^d}
\end{equation}
for $d \ll n$. The target mapping $\xi(\bx)$ can be linear, nonlinear, or even an identity function (i.e., this reduces the problem to selection). \req{eq:collective-variables} is central to our review: each manifold learning method provides a unique functional form of the target mapping used to reduce the dimensionality of the system representation.

To define a probability density for CVs expressed by the target mapping (\req{eq:collective-variables}), we need to consider only a part of the configuration space. The equilibrium distribution of CVs is obtained by averaging over unused variables. This gives us a marginal density:
\begin{align}
  \label{eq:collective-variables-density}
  \rho(\bz) = \int\dx \delta\ppar*{\bz - \xi(\bx)} \rho(\bx)
         = \pang[\Big]{\delta\ppar*{\bz - \xi(\bx)}},
\end{align}
where the multidimensional Dirac delta function is:
\begin{equation}
  \delta\ppar{\bz-\xi(\bx)}=\prod_{k=1}^d \delta\ppar{z_k-\xi_k(\bx)},
\end{equation}
and $\pang*{\cdot}$ denotes an unbiased ensemble average. The equilibrium distribution of CVs (\req{eq:collective-variables-density}) typically contains several disconnected states of high probability separated by regions of low probability leading to infrequent transitions between such states.

\subsection{Free-Energy Landscape}
\label{sec:free-energy}
\begin{figure}
  \includegraphics{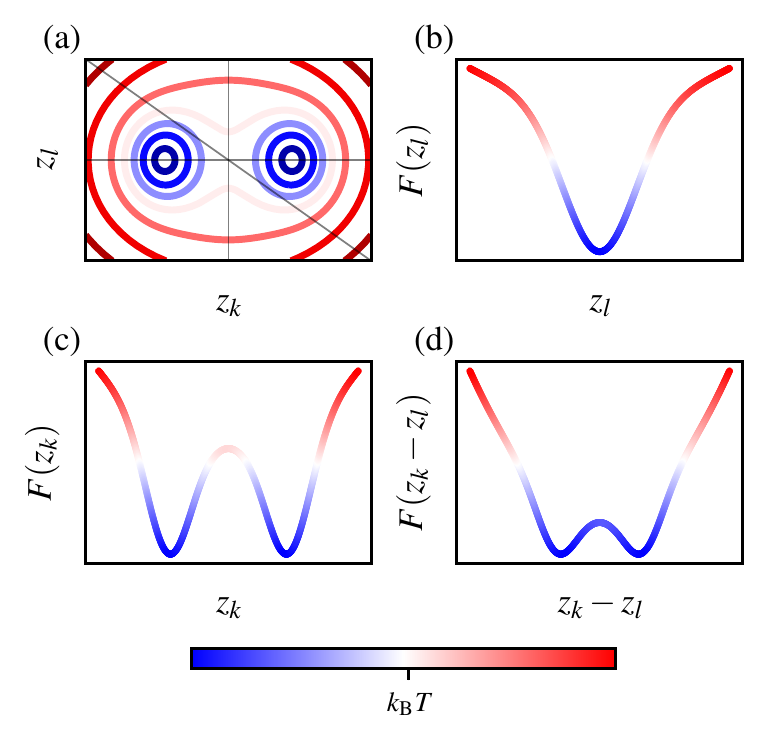}
  \caption{{\bf Metastability}.
  \chg{(a) Simplified model of a system with two long-lived metastable states and its free-energy landscape $F(z_k,z_l)$, where $\bz=(z_k,z_l)$, with a free-energy barrier higher than the thermal energy $\kT$. Gray lines show selected variables for projections shown in (b-d) with remaining variables integrated out, e.g., for $\bz=(z_k,z_l)$, $F(z_k)=-\frac{1}{\beta}\log\int\mathrm{d}z_l\,\e^{-\beta F(z_l)}$.}
  (b) Free energy along the $z_l$ variable indicates only one minimum, which means that $z_l$ is not an optimal CV.
  (c) Free energy along $z_k$ depicts two minima with a correct free-energy barrier, slightly higher than the thermal energy (see color bar).
  (d) Free energy along $z_k-z_l$ shows two energy minima, but the free-energy barrier is lower than the correct value shown in (a).}
  \label{fig:metastable-potential}
\end{figure}

Instead of the potential energy function $U(\bx)$ characteristic for a high-dimensional representation (\req{eq:boltzmann-density}), the reduced dynamics of the system in the CV space follows the underlying \emph{free-energy landscape}. We define it as the negative logarithm of the marginal distribution of CVs multiplied by the thermal energy:
\begin{equation}
  \label{eq:free-energy}
  \boxed{%
    F(\bz) = -\frac{1}{\beta}\log \rho(\bz)}
\end{equation}
which is defined up to an immaterial constant. The equilibrium density of CVs can be equivalently written as $\rho(\bz) = \e^{-\beta F(\bz)} / \mathcal{Z}$, where the partition function in the CV space is given as $\mathcal{Z}=\int\dz\e^{-\beta F(\bz)}$. 

The free-energy landscape determines an effective energy landscape for CVs that consists of kinetic barriers between metastable states. The free-energy difference between states $A$ and $B$ can be calculated as~\cite{henin2022enhanced}:
\begin{equation}
  \label{eq:free-energy-difference}
  \Delta F_{AB} = -\frac{1}{\beta} \log{ \frac{\mathcal{Z}_{A}}{\mathcal{Z}_{B}}} = -\frac{1}{\beta} \log{\frac{\int_A\dz\, \e^{-\beta F(\bz)}}{\int_B\dz\, \e^{-\beta F(\bz)}}},
\end{equation}
which is defined using the ratio between the partition functions corresponding to the states $\mathcal{Z}_A$ and $\mathcal{Z}_B$, respectively. Note that \req{eq:free-energy-difference} is valid if (and only if) the CV set properly separates the two states $A$ and $B$. To calculate the free-energy difference, one must integrate over the regions in CV space defining the states $A$ and $B$, as simply taking the difference of $F(\bz)$ at the minima of states $A$ and $B$ does not yield correct results~\cite{dietschreit2022obtain}. 

\chg{  
As the free-energy landscape (\req{eq:free-energy}) is not invariant with respect to CVs~\cite{bal2020free,dietschreit2022obtain,dietschreit2022free}, the relation of the free-energy barrier to the kinetics of crossing between states~\cite{bal2020free} or activation free energies~\cite{dietschreit2022free} is not apparent (\rfig{fig:metastable-potential}). 

Sampling free-energy landscapes exhaustively can be challenging, even for simple systems. On the timescales accessible for standard atomistic simulations (around milliseconds), crossings over high free-energy barriers are {\it rare events}. As a result, the system remains kinetically trapped in a metastable state as its dynamics is restricted to sampling fast equilibration. This so-called \emph{sampling problem} can be observed in many physical processes, for example, catalysis~\cite{piccini2022ab}, ligand interactions with proteins~\cite{baron2013molecular,rydzewski2017ligand,bruce2018new,bernetti2019kinetics,wolf2023predicting} and DNA~\cite{o2021enhanced}, glass transitions in amorphous materials~\cite{van2021towards}, crystallization~\cite{neha2022collective}, and graphite etching~\cite{aussems2017atomistic}.
}

\subsection{Enhanced Sampling}
\label{sec:enhanced-sampling}
To alleviate the sampling problem and overcome kinetic bottlenecks, enhanced sampling methods can be used. Over recent years, several such enhanced sampling algorithms have been developed, including tempering~\cite{parallel_tempering,earl2005parallel,chen2012heating}, variational~\cite{valsson2014variational,reinhardt2020determining}, biasing~\cite{torrie1977nonphysical,mezei1987adaptive,laio2002escaping,barducci2008well,Maragakis-JPCB-2009,morishita2012free} approaches, or combinations of these~\cite{invernizzi2020unified}. For a comprehensive review and classification of these methods, we refer to an article by Henin et al.~\cite{henin2022enhanced}.

\chg{
As a representative example of enhanced sampling techniques, we consider methods that employ an external bias potential to enhance CV fluctuations. The first approach of this kind, called umbrella sampling, was introduced in 1977 by Torrie and Valleau~\cite{torrie1977nonphysical}. The motivation for naming the method ``umbrella sampling'' was to highlight the versatility of the method to investigate a wide range of physical processes~\cite{battimelli2020computer}.
}

When the bias potential is introduced to the system, the distribution of CVs can deviate significantly from equilibrium (\req{eq:collective-variables-density}). This results in sampling according to a biased distribution:
\begin{align}
  \label{eq:collective-variables-density-biased}
  \rho_V(\bz) = \frac{1}{\mathcal{Z}_V} \e^{-\beta \ppar*{F(\bz) + V(\bz)}}
           = \pang[\Big]{\delta\ppar*{\bz - \xi(\bx)}}_V,
\end{align}
where $\mathcal{Z}_V = \int\dz\e^{-\beta \ppar*{F(\bz) + V(\bz)}}$ is the biased partition function and $\pang*{\cdot}_V$ denotes an ensemble average calculated under the biasing potential $V(\bz)$ . By design, the biased distribution is easier to sample.

Enhanced sampling methods based on CVs vary in how they flatten or reduce free-energy barriers and how they construct the bias potential~\cite{torrie1977nonphysical,laio2002escaping,barducci2008well,Invernizzi2020opus,valsson2014variational}. In umbrella sampling, it was proposed that the biased probability distribution of CVs should be ``as wide and uniform as possible''~\cite{torrie1977nonphysical}, and thus sometimes called flat histogram. However, recent developments in enhanced sampling have shown that this approach may not be the most efficient for fast convergence~\cite{dayal2004performance,trebst2004optimizing,barducci2008well,valsson2015well,invernizzi2020unified}.

\begin{figure*}
  \includegraphics{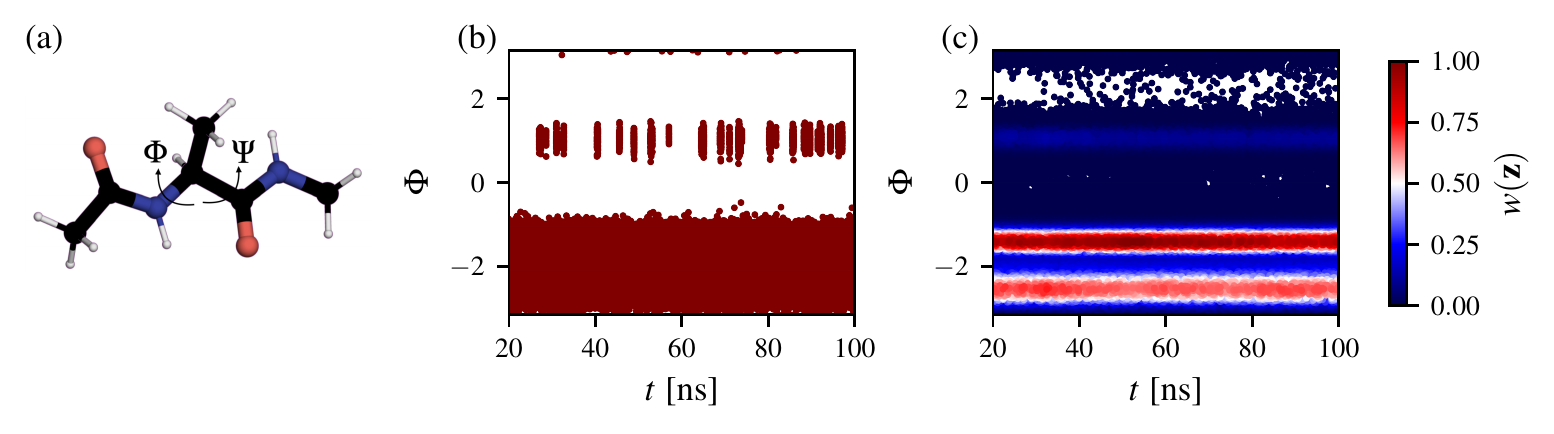}
  \caption{{\bf Enhanced Sampling and Statistical Weights}. (a) Alanine dipeptide in vacuum and its dihedral angles $\Phi$ and $\Psi$. (b) Sampling of the $\Phi$ dihedral angle of alanine dipeptide performed using a parallel tempering simulation. The timeseries shows the replica at 300 K. All the samples are equally important as they are sampled from the equilibrium distribution. (c) Enhanced sampling of $\Phi$ performed using well-tempered metadynamics at 300 K with a bias factor $\gamma$ of 5. Color corresponds to statistical weights of the samples $w(\bz)$, which vary considerably with the most important samples belonging to the metastable states.}
  \label{fig:md}
\end{figure*}

As an example of a non-uniform target distribution, let us consider the well-tempered distribution used in metadynamics~\cite{barducci2008well}. Well-tempered metadynamics uses a history-dependent bias potential updated iteratively by depositing Gaussians centered at the current location in the CV space:
\begin{equation}
  \label{eq:bias-potential}
  V(\bz) = \sum_k G_{\sigma}(\bz,\bz_k) \exp\ppar*{-\frac{1}{\gamma-1}\beta V(\bz_k)},
\end{equation}
where $G_{\sigma}(\bz,\bz_k)$ is a Gaussian kernel with a bandwidth set $\sigma$, $\bz_k$ is the center of $k$-th added Gaussian, and $\gamma$ is a bias factor that determines how much we enhance CV fluctuations. Well-tempered metadynamics convergences to the well-tempered distribution:
\begin{equation}
  \label{eq:wt-distribution}
  \rho_V(\bz) \propto \pbkt*{ \rho(\bz) }^{1/\gamma},
\end{equation}
in which we sample an effective free-energy landscape $F_{\gamma}(\bz)=F(\bz)/\gamma$ with barriers reduced by a factor of $\gamma$~\cite{barducci2008well,henin2022enhanced}.

\subsection{Reweighting Biased Probability Distributions}
\label{sec:reweighting}
When conducting biased simulations, the CVs sample a smoother biased CV distribution. When obtaining equilibrium properties, such as free-energy landscapes, each sample is given a statistical weight to account for the effect of the biasing. For methods employing a quasi-stationary bias potential $V(\bz)$, the weight associated with a CV sample $\bz$ is given as:
\begin{equation}
  \label{eq:weight}
  \boxed{%
    w(\bz) = \e^{\beta V(\bz)}}.
\end{equation}
In contrast, well-tempered metadynamics uses an adaptive bias potential (\req{eq:bias-potential}), and the weights are modified by adding a time-dependent constant to the bias potential~\cite{tiwary_rewt,valsson2016enhancing}. In unbiased simulations, every sample is equally important as it is obtained from the equilibrium distribution (\rfig{fig:md}). 

Standard reweighting involves employing the weights to find the stationary equilibrium distribution from the biased CV distribution, i.e., $\rho(\bz) \propto w(\bz) \rho_V(\bz)$, which can be computed by histogramming or kernel density estimation, where each sample $\bz$ is weighted by \req{eq:weight}. Many simulation codes, such as PLUMED~\cite{plumed,plumed-nest}, routinely use this approach.

\section{Simulation Data}
\label{sec:simulation-data}
Here, we describe how to prepare a dataset from standard atomistic and enhanced sampling simulations for a manifold learning method. These data have unique characteristics that differ from ordinary datasets. Moreover, we demonstrate how to collect such datasets and reduce their size while preserving their density.

\subsection{Data Representation}
\chg{To prepare a simulation dataset, we collect high-dimensional samples represented by the configuration variables. This dataset can be a trajectory of the system, as shown in \req{eq:trajectory}. Additionally, we can sample the configuration variables in multiple trajectories that originate from the same probability distribution and combine them. If the dataset contains high-dimensional samples from the equilibrium probability distribution, we do not require additional information about the system.

However, as explained in \rsct{sec:reweighting}, if we sample simulation data from a biased probability density, each high-dimensional sample has a statistical weight that contains information about its importance. Therefore, we can express that dataset as:
\begin{align}
  \label{eq:biased-data}
  \boxed{%
    X = \pset[\big]{\ppar*{\bx_k \in \mathbb{R}^n, w(\bx_k)}}_{k=1}^K},
\end{align}
where the weights are given by (\req{eq:weight}). Failure to consider these weights when creating a dataset from biased simulations data can affect its geometry, density, and importance~\cite{rydzewski2021multiscale,rydzewski2022reweighted,rydzewski2023selecting}.
}

\chg{
A fundamental issue of simulating complex systems is the convergence and accuracy of the data. When the simulation fails to capture enough transitions between long-lived metastable states, the representation of these rare events is greatly affected. This problem is especially pronounced in systems exhibiting metastability and sampling problems. For useful convergence measures in atomistic simulations, see~\rref{hess2002convergence,romo2011block,grossfield2018best}.

It is preferable for the dataset to have uncorrelated samples. However, simulations often generate correlated samples on shorter timescales. To avoid correlation, samples from simulations can be collected with a large enough time stride. For a detailed discussion about correlation in atomistic simulations, see~\rref{grossfield2018best}.
}

\subsection{Landmark Sampling}
\label{sec:landmark-sampling}
\chg{
Vast amounts of data obtained from atomistic simulations often necessitate reducing the number of samples in the datasets. This can be achieved through \emph{landmark sampling}, where a subset of the samples from the high-dimensional dataset is selected to create a dataset with preferably much fewer samples that retain information about the complete dataset. This approach is used both machine learning, where low-rank approximations such as the Nystr\"om extensions and clustering are widely used~\cite{bengio2003out,de2003global,de2004sparse,silva2005selecting,belabbas2009landmark,aflalo2013spectral,hong2023two}, and atomistic simulations~\cite{das2006low,ceriotti2013demonstrating,zhang2018unfolding,long2019landmark,kahle2019unsupervised,rydzewski2021multiscale,vymetal2022iterative}. Below, we focus on landmark sampling techniques that can be modified to include statistical weights from enhanced sampling simulations as a selection criterium.
}

When working with unbiased data, it is reasonable to choose landmarks randomly, as all samples are equally important (\req{eq:biased-data}). This approach results in landmarks that are approximately distributed according to the Boltzmann equilibrium distribution. Farthest point sampling (FPS)~\cite{hochbaum1985best} can also be used to sample landmarks from unbiased simulations. FPS relies on geometric criteria to ensure a uniform selection of landmarks, regardless of the probability distribution. Concretely, FPS selects a sample that is farthest from all previously selected landmarks. However, FPS may be computationally expensive for large datasets.

For biased trajectories, landmark samples can be chosen based on their weights $w(\bx)$ (\req{eq:biased-data}). In the simplest case, landmarks can be obtained by \emph{weighted random sampling}, where each sample is selected with a probability proportional to its weight~\cite{bortz1975new}. However, this method can result in an overpopulation of samples in the deepest free-energy basins while leaving other metastable states with a limited number of samples~\cite{ceriotti2013demonstrating,tribello2019using_a,tribello2019using_b,rydzewski2021multiscale}. This overpopulation occurs because samples in the deepest metastable states have much higher weights than those in higher-lying states due to the exponential dependence of the statistical weights on the bias potential.

To exploit the information about the geometry and density of the configuration space, we can employ a well-tempered variant of FPS~\cite{ceriotti2013demonstrating}. This selection process involves two stages. First, FPS selects $\sqrt{KL}$ samples, which are then divided into Voronoi polyhedra $\pset{v_k}$. For each polyhedron $v_k$, we calculate a tempered weight: 
\begin{equation}
  \label{eq:acc-weights}
  \omega_k(\bx) = \ppar*{\sum_l w(\bx_l)}^{1/\alpha},
\end{equation}
where $l \in v_k$ and $\alpha>1$ is a tempering parameter. In the second stage, weighted random sampling selects a polyhedron according to the tempered weight (\req{eq:acc-weights}). A landmark is then sampled from the selected Voronoi polyhedron using unmodified weights (without tempering). This process is repeated until the desired number of landmarks $L$ is reached. In the limit of $\alpha\rightarrow\infty$, landmarks are uniformly distributed. For $\alpha\rightarrow 1$, the landmarks selection should resemble the equilibrium distribution. The procedure for well-tempered FPS is summarized in Algorithm~\ref{alg:wt-fps}.
\begin{algorithm}
  \label{alg:wt-fps}
  \SetKwInOut{Input}{Input}
  \SetKwInOut{Output}{Output}
  \Input{Biased data batch $X=\pset[\big]{(\bx_k, w(\bx_k))}_{k=1}^K$; \chg{tempering parameter $\alpha$}; number of landmarks $L$.}
  \Output{Training set of landmarks $\pset{\bx_k}_{k=1}^L \in X$.}
  \begin{enumerate}[leftmargin=0cm]
    \item Select $\sqrt{KL}$ landmarks using FPS.
    \item Perform the Voronoi tesselation $\pset{v_k}$ based on landmarks selected using FPS.
    \item Calculate the accumulated weights for each Voronoi polyhedra $\omega_k(\bx)$ for $\alpha$ (\req{eq:acc-weights}).
    \item Until the number of landmarks is $L$:
    \begin{enumerate}[leftmargin=0.6cm]
      \item Select a Voronoi polyhedra according to the accumulated weights $\omega_k$.
      \item Include a landmark based on weighted random sampling.
    \end{enumerate}
  \end{enumerate}
  \caption{Well-tempered FPS.}
\end{algorithm}

Recently, weight-tempered random sampling (WTRS) has been introduced as an alternative for well-tempered FPS~\cite{rydzewski2021multiscale}. This technique involves scaling the weights of samples without using FPS. WTRS selects a landmark with a probability $\propto w^{1/\alpha}$, where, as in \req{eq:acc-weights}, $\alpha$ is a tempering parameter. Assuming landmarks are sampled from the well-tempered distribution (\req{eq:wt-distribution}), the limit $\alpha \rightarrow \infty$ corresponds to sampling landmarks according to the biased distribution, while the limit $\alpha \rightarrow 1$ results in sampling landmarks from the equilibrium distribution~\cite{rydzewski2021multiscale}.
\begin{figure}
  \includegraphics{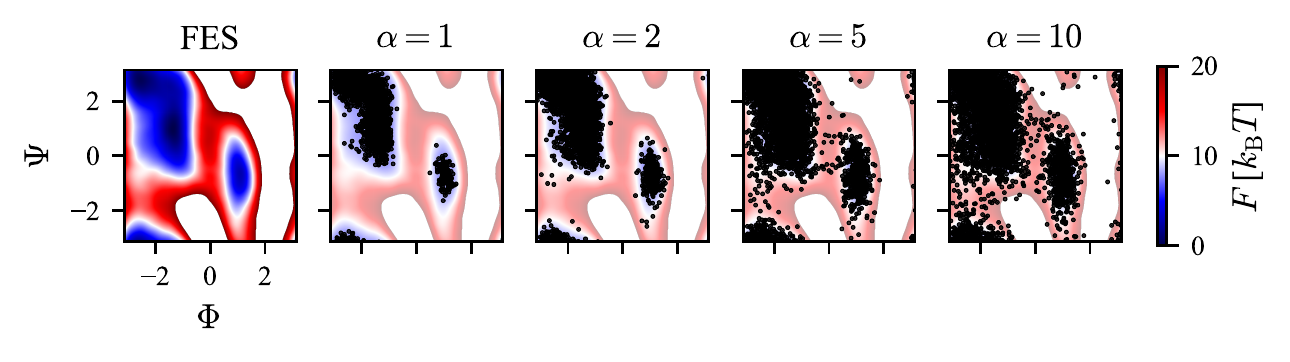}
  \caption{{\bf Landmark Sampling with Weight-Tempered Random Sampling}. Sampling landmarks according to probabilities $\propto w^{1/\alpha}$ for different values of the tempering parameter $\alpha$ compared to the free-energy surface (FES) of alanine dipeptide in vacuum. Biased simulation data is generated from a well-tempered metadynamics simulation of alanine dipeptide in vacuum using a bias factor of $\gamma=5$ and the $\Phi$ and $\Psi$ dihedral angles as biased variables. By increasing $\alpha$, landmarks gradually diverge from the unbiased distribution. Data taken from~\rref{rydzewski2021multiscale}.}
\end{figure}

\section{Manifold Learning}
\label{sec:manifold-learning}
In this section, we discuss fundamental concepts of manifold learning techniques. We aim to explain how these methods can be employed to analyze simulation data and construct CVs. A non-exhaustive list of manifold learning methods that can be applied to atomistic systems includes locally linear embedding~\cite{roweis2000nonlinear}, Laplacian eigenmap~\cite{belkin2001laplacian,belkin2003laplacian,bengio2004learning}, diffusion map~\cite{coifman2005geometric,coifman2006diffusion,nadler2006diffusion,coifman2008diffusion}, spectral gap optimization of order parameters~\cite{tiwary2016spectral}, sketch-map~\cite{ceriotti2011simplifying,ceriotti2013demonstrating}, and stochastic neighbor embedding and its variants~\cite{hinton2002stochastic,maaten2008visualizing,maaten2009learning}.
\subsection{Markov Transition Matrix}
\label{sec:markov-matrix}

\chg{Manifold learning, also referred to as nonlinear dimensionality reduction, aims to simplify high-dimensional data to low-dimensional manifolds. This class of techniques generalizes linear methods such as principal component analysis (PCA) or singular value decomposition (SVD). Manifold learning is commonly used to analyze machine-learning data and has recently gained attention in atomistic simulations for extracting physical properties from the dynamics of complex systems. In the following, we consider a manifold as a low-dimensional space spanned by a few CVs.}

\chg{
A primary assumption in manifold learning for atomistic systems is that the dynamics in a high-dimensional space can be represented by a low-dimensional and smooth subspace called a manifold. This assumption is known as the \emph{manifold hypothesis}. It states that the dynamics effectively evolves on the low-dimensional manifold embedded in the high-dimensional space~\cite{ferguson2011integrating,ferguson2011nonlinear}. The existence of such low-dimensional descriptions may be attributed to couplings between the degrees of freedom, resulting in a small number of slowly evolving variables that govern the dynamics and to which the remaining fast degrees of freedom (or their statistics) are slaved. Under this view, the fast degrees of freedom are controlled by the dynamics of the slow CVs due to fast equilibration within the metastable states, resulting in an adiabatic timescale separation. This description leads naturally to the modeling of complex systems as diffusion processes, in which a set of stochastic differential equations may be formulated in the slow variables, with the fast degrees of freedom represented as thermal noise. Thus, characterizing the system dynamics by the slow CVs leads to a negligible error. 
}

\chg{The core of most manifold learning methods is having a notion of similarity between high-dimensional samples, usually through a distance metric~\cite{roweis2000nonlinear,tenenbaum2000global,belkin2001laplacian,hinton2002stochastic,belkin2003laplacian,hashemian2013modeling,coifman2005geometric,maaten2008visualizing,hashemian2013modeling,aflalo2013spectral}. The distances are then integrated into a global parameterization of the data through the construction of a discrete Markov chain, where the similarities depend on distances between the samples. For example, a common starting point is the construction of the Markov chain based on a Gaussian kernel:
\begin{equation}
  \label{eq:gaussian-kernel}
  G_\varepsilon(\bx_k,\bx_l) = \exp\ppar*{-\frac{1}{\varepsilon} \| \bx_k - \bx_l\|^2},
\end{equation}
where $\|\cdot\|$ denotes Euclidean distances. Manifold learning techniques discussed in this review use a normalized Gaussian kernel with Euclidean distances to model the Markov chain. The distances can be computed between all samples in the dataset or only between nearest neighbors. The scale parameter $\varepsilon>0$ depends on the dataset as it induces a length scale ${\sim}\sqrt{\varepsilon}$. It can be selected to match the distance between neighboring samples. We cover different algorithms for selecting $\varepsilon$ when discussing each method (\rsct{sec:tm-eigendecomposition} and \ref{sec:tm-divergence-optimization}).

Depending on distance metrics, other Gaussian kernels can be considered~\cite{ham2004kernel}, for instance, generalized Minkowski, Mahalanobis, or cosine distances. Moreover, Laplacian~\cite{belkin2001laplacian,belkin2003laplacian}, heat kernel~\cite{berard1994embedding,jones2008manifold}, weight matrix~\cite{roweis2000nonlinear}, graph diffusion kernel~\cite{kondor2002diffusion}, or Fisher information kernel~\cite{lafferty2005diffusion} can also be used in manifold learning. Any kernel $G(\bx_k,\bx_l)$ used in the construction of the Markov chain must satisfy the following properties:
\begin{enumerate}[leftmargin=0.5cm]
  \item $G$ is symmetric: $G(\bx_k,\bx_l)=G(\bx_l,\bx_k)$.
  \item $G$ is positivity preserving: for all samples $G(\bx_k,\bx_l) \ge 0$.
  \item $G$ is positive semi-definite.
\end{enumerate}

To convert the Gaussian kernel into a Markov transition matrix $M(\bx_k,\bx_l)$ of size $K\times K$ that describes transition probabilities $p_{kl}$, we row-normalize \req{eq:gaussian-kernel} to obtain a right stochastic matrix:
\begin{equation}
\label{eq:gaussian-markov-matrix}
  p_{kl} \sim M(\bx_k,\bx_l) = \frac{G_\varepsilon(\bx_k,\bx_l)}{\sum_n G_\varepsilon(\bx_k,\bx_n)},
\end{equation}
which describes the probability of transitioning from sample $\bx_k$ to sample $\bx_l$ in an auxiliary time step $t$:
\begin{align}
\label{eq:markov-chain}
   M(\bx_k, \bx_l) = \mathrm{Pr}\,\pset[\big]{\bx_{t+1}=\bx_l\,|\,\bx_{t}=\bx_k},
\end{align}
where, through the construction in \req{eq:gaussian-markov-matrix}, the transition probability depends only on the current sample, i.e., it is called the \emph{Markovian} assumption, meaning that the system has no memory. 
}

The Markov transition matrix can be used to propagate the corresponding Markov chain. For a time-homogeneous Markov chain, the $k$-step transition probability can be obtained as the $k$-th power of the transition matrix $M^k$. A unique stationary distribution $\pi(\bx)$ exists if the Markov chain is irreducible and aperiodic. In this case, $M^k$ converges to a rank-one matrix in which each row is the stationary distribution $\lim_{k\rightarrow\infty} M^k = \pi(\bx)$. Alternatively, it can be as defined $M\pi(\bx)=\pi(\bx)$ as the stationary distribution is unchanged by the Markov transition matrix. Consequently, the highest eigenvalue corresponding to the stationary distribution equals one.

The terminology used to describe $M(\bx_k,\bx_l)$ can vary depending on the field from which a method originates. In unsupervised learning, it is commonly referred to as the \emph{affinity}, \emph{similarity}, or \emph{proximity} matrix~\cite{hinton2002stochastic,ham2004kernel,maaten2008visualizing,mcinnes2018umap}. These methods usually include an additional step when building a Markov chain, e.g., the diagonal entries of $M$ are set to zero. This type of Markov chain is called \emph{non-lazy}. In contrast, methods devised for atomistic systems do not use this assumption as the diagonal entries may contain important information about the system~\cite{nadler2006diffusion}. In such Markov chains, there is a probability that the transition does not occur.

\begin{figure}
  \includegraphics{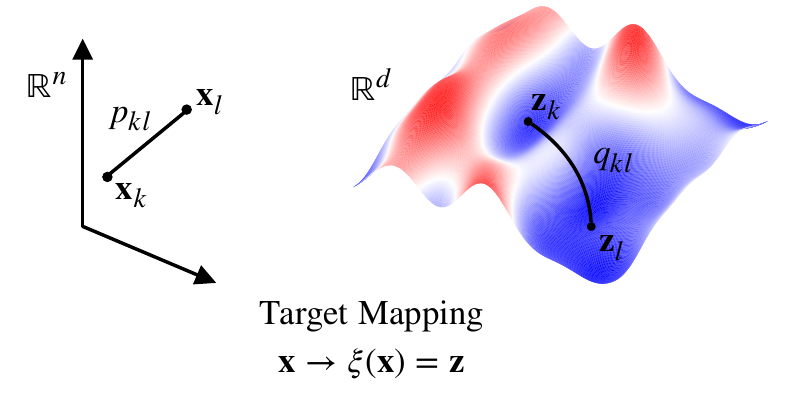}
  \caption{{\bf Learning Manifolds and Target Mapping}.
  Schematic representation of embedding high-dimensional samples $\pset{\bx}$ into a low-dimensional manifold. The target mapping $\xi(\bx): \mathbb{R}^n \mapsto \mathbb{R}^d$ is defined such that relations $p_{kl}$ between high-dimensional samples $\bx_k$ and $\bx_l$ is preserving relations $q_{kl}$ between low-dimensional samples $\bz_k$ and $\bz_l$ in the manifold. Learning a manifold is equivalent to learning a mapping which can be constructed using eigendecomposition (\rsct{sec:tm-eigendecomposition}) or divergence optimization (\rsct{sec:tm-divergence-optimization}).}
  \label{fig:manifold}
\end{figure}

\subsection{Target Mapping}
Here, we focus on a generalization of the target mapping from the high-dimensional configuration space (\req{eq:configuration-variable}) to a low-dimensional manifold (or the CV space) (\req{eq:collective-variables}). As explained in \rsct{sec:collective-variables}, under our framework, the problem of finding CVs is equivalent to finding a parametrization of the target mapping. The target mapping performs dimensionality reduction such that the dimensionality of the manifold is much lower than that of the high-dimensional space, i.e., $d \ll n$; see \rfig{fig:manifold}.

Manifold learning methods can be split into two categories depending on how the low-dimensional manifold is constructed by the target mapping:
\begin{enumerate}[leftmargin=0.5cm]
  \item[I.] \emph{Eigendecomposition} (i.e., spectral decomposition) of the Markov transition matrix:
  \begin{equation}
    \label{eq:target-mapping-eigen}
    M \psi_k = \lambda_k \psi_k,
  \end{equation}
  where $\pset{ \psi_k }$ and $\pset{ \lambda_k }$ are the corresponding eigenfunctions and eigenvalues, respectively. The solution of \req{eq:target-mapping-eigen} determines the low-dimensional manifold~\cite{coifman2005geometric}. For instance, the target mapping can be parametrized as follows:
  \begin{equation}
    \xi(\bx) = \pset[\big]{ \lambda_k \psi_k(\bx) }_{k=1}^d,
  \end{equation}
  where the $k$-th manifold coordinate is $\lambda_k \psi_k(\bx)$. The eigenvalues are sorted in non-increasing order and include only $d$ dominant eigenvalues as each corresponds to the importance of respective coordinates spanned by eigenfunctions.

  The eigenvalues decrease exponentially and can be related to the effective timescales of the studied physical process, as multiple timescales frequently characterize complex systems. As such, the dominant eigenvalues also correspond to the slowest processes. We analyze this in detail in \rsct{sec:diffusion-map}. Manifold learning methods exploiting the eigendecomposition to find a low-dimensional manifold of CVs are described in \rsct{sec:tm-eigendecomposition}.

  \item[II.] \emph{Divergence optimization} where a divergence (i.e., a statistical distance between a pair of probability distributions) between the Markov transition matrix $M$ built from high-dimensional samples and a Markov transition matrix $Q(\bz_k,\bz_l)$, constructed from low-dimensional samples, is minimized (\rfig{fig:manifold}). The target mapping expressed as a parametrizable embedding is:
  \begin{equation}
    \label{eq:div-opt-mapping}
    \xi_{\btheta}(\bx) = \pset[\big]{ \xi_k(\bx;\btheta) }_{k=1}^d,
  \end{equation}
  where $\btheta=\pset{\theta_k}$ are parameters that are varied such that the divergence between $M$ and $Q$ is minimized. In such methods, $M$ is fixed while $Q$ is estimated by the parametrized target mapping. Depending on the manifold learning method used, the minimization can be performed differently, i.e., gradient descent~\cite{maaten2008visualizing}, or stochastic gradient descent if the target mapping is represented by a neural network~\cite{maaten2009learning,zhang2018unfolding,rydzewski2021multiscale}. We introduce such manifold learning methods in \rsct{sec:tm-divergence-optimization}.
\end{enumerate}

\section{Target Mapping (I): Eigendecomposition}
\label{sec:tm-eigendecomposition}
Here, we cover manifold learning methods that employ the eigendecomposition of the Markov transition matrix to find the target mapping $\xi(\bx)$. The following manifold learning methods we discuss are diffusion map (DMAP) (\rsct{sec:diffusion-map}), time-lagged independent component analysis (TICA) (\rsct{sec:tica}), and spectral gap optimization (SGOOP) (\rsct{sec:spectral-gap-optimization}).

\subsection{Diffusion Map (DMAP)}
\label{sec:diffusion-map}
The concept of DMAP was inspired mainly by Laplacian eigenmap. Laplacian eigenmap~\cite{belkin2001laplacian,belkin2003laplacian} originates from spectral graph theory~\cite{chung1997spectral}, which is a mathematical field studying properties of the Laplacian matrix or adjacency matrix associated with a graph. As states of the system can be represented as a graph, Laplacian eigenmap is commonly used in manifold learning to reduce data dimensionality. Laplacian eigenmap has theoretical convergence guarantees as the discrete operator approaches the Laplacian on the underlying manifold assuming the data are \textit{uniformly} sampled.

DMAP proposed by Coifman et al.~\cite{coifman2005geometric} expands the concept of Laplacian eigenmap. This algorithm yields a family of embeddings and provides theoretical understanding for the resulting embedding even when the data are \textit{non-uniformly} sampled. DMAP can construct an informative low-dimensional embedding of a complex system. Compared to other manifold learning methods, DMAP has a substantial theoretical background~\cite{coifman2005geometric,coifman2006diffusion,coifman2008diffusion}. Namely, the CVs spanning the low-dimensional manifold constructed by DMAP correspond to the slowest relaxation processes given by a probability distribution evolving under a random walk over the data~\cite{nadler2006diffusion}.

\chg{
Many variants of DMAP have been developed for finding manifolds of physical systems, for instance, anisotropic~\cite{coifman2008diffusion,singer2009detecting}, locally scalled~\cite{rohrdanz2011determination}, variationally optimized~\cite{boninsegna2015investigating}, target measure~\cite{banisch2020diffusion}, and reweighted~\cite{rydzewski2022reweighted,rydzewski2023selecting}. 
}

\subsubsection{Anisotropic Diffusion Kernel}
\label{sec:diffusion-kernel}
\chg{  
As a starting point for our framework, we consider the anisotropic DMAP~\cite{nadler2006diffusion}. DMAP first employs a Gaussian kernel function to estimate the similarity between high-dimensional feature samples $\bx_k$ and $\bx_l$, $G_\varepsilon(\bx_k,\bx_l)$, where $\varepsilon$ is a scale parameter (\req{eq:gaussian-kernel}). The scale parameter in DMAP $\varepsilon$ can be chosen using several heuristics. For instance, see \rref{kim2015systematic,berry2016variable,lindenbaum2020gaussian}.
}

Next, the Gaussian kernel is used to define a density-preserving kernel in a high-dimensional space that is called the anisotropic diffusion kernel:
\begin{equation}
  \label{eq:diffusion-map-kernel}
  L(\bx_k,\bx_l) = \frac{G_\varepsilon(\bx_k,\bx_l)}{[\varrho(\bx_k)]^\alpha [\varrho(\bx_l)]^\alpha},
\end{equation}
where $\varrho(\bx_k) = \sum_n G_\varepsilon(\bx_k,\bx_n)$ is up to a normalization constant a pointwise kernel density estimate at $\bx_k$ and $\alpha \in [0,1]$ is a normalization parameter, the anisotropic diffusion constant, based on which a family of different manifold parametrizations can be considered~\cite{coifman2005geometric}.

Then, \req{eq:diffusion-map-kernel} is row-normalized to represent Markov transition probabilities:
\begin{equation}
  \label{eq:diffusion-map-markov}
  p_{kl} \sim M(\bx_k,\bx_l) = \frac{L(\bx_k,\bx_l)}{\sum_n L(\bx_k,\bx_n)},
\end{equation}
so that $\sum_l M(\bx_k,\bx_l) = 1$ for $k$-th row. As such, \req{eq:diffusion-map-kernel} is a Markov transition matrix containing information about the transition probability from $\bx_k$ to $\bx_l$. Under this view, \req{eq:diffusion-map-markov} denotes a Markov chain with the transition probability from $\bx_k$ to $\bx_l$ in an auxiliary time step (\req{eq:markov-chain}). It can be shown that there is a direct link between Laplacian eigenmap and DMAP~\cite{nadler2006diffusion}.

The anisotropic diffusion constant (\req{eq:diffusion-map-kernel}) is related to the importance of data density~\cite{coifman2006diffusion}. Specifically, when the microscopic coordinates are sampled according to the equilibrium density, in the limits $\varepsilon\rightarrow 0$ and $K\rightarrow\infty$, DMAP asymptotically converges to the stationary Boltzmann distribution of the modeled Markov chain. Based on these assumptions and depending on the normalization using the anisotropic diffusion constant, we can consider the following interesting constructions of the stationary density:\chg{
\begin{enumerate}[leftmargin=0.5cm]
  \item $\alpha=0$: we recover dynamics according to the potential $2 U(\bx)$ and the density $\propto \pbkt{\rho(\bx)}^2$ for the classical normalized graph Laplacian~\cite{belkin2001laplacian,belkin2003laplacian,jones2008manifold}.
  \item $\alpha=1$: we get the graph Laplacian with data uniformly distributed on a manifold. This normalization accounts only for the data geometry, while density does not play a role.
  \item $\alpha=\frac{1}{2}$: we obtain dynamics according to the underlying potential $U(\bx)$ and the density $\propto \rho(\bx)$ whose eigenfunctions capture the long-time asymptotics of data (i.e., correspond to slow variables).
\end{enumerate}
}

For the anisotropic diffusion constant $\alpha=\frac{1}{2}$, we asymptotically recover the long-time dynamics of the system whose microscopic coordinates are sampled from the Boltzmann distribution (\req{eq:boltzmann-density}). The related backward Fokker--Planck differential equation is given byy~\cite{coifman2006diffusion}:
\begin{equation}
  \label{eq:sde}
  \mu \dx = -\nabla U(\bx)\mathrm{d}t + \sqrt{2\mu{\beta}^{-1}}\dd{\mathbf{w}},
\end{equation}
where $\mu$ is the friction coefficient, $-\nabla U(\bx)$ denotes the force acting on atoms, and $\mathbf{w}$ is an $n$-dimensional Brownian motion. The infinitesimal generator of this diffusion process is:
\begin{equation}
  \label{eq:infgen}
  \mathcal{L}=-\mu^{-1} \nabla U \cdot \nabla + (\mu\beta)^{-1} \nabla^2.
\end{equation}
The eigenvalues and eigenvectors of $\mathcal{L}$ determine the kinetic information of the diffusion process and can be used to parametrize a low-dimensional manifold.

As we are interested in finding a low-dimensional representation of a system, i.e., estimating CVs, the case for $\alpha=\frac{1}{2}$ is crucial to model the slowest degrees of freedom, accounting for both the underlying geometry and density of the manifold.

Note that when considering configuration variables other than the microscopic coordinates, the underlying equilibrium density is not given by the Boltzmann distribution (\rsct{sec:atomistic-simulations}). However, it is reasonable to assume that there is a separation of timescales for variables other than the microscopic coordinates.

\subsubsection{Diffusion Coordinates}
The idea of using eigenfunctions of the Gaussian kernel as coordinates for Riemannian manifolds originates with \rref{berard1994embedding} and in the context of data analysis with \rref{coifman2006diffusion}. The transition probability matrix $M$ can be used to solve the eigenvalue problem:
\begin{equation}
  M\psi_k = \lambda_k \psi_k
\end{equation}
for $k=1, \dots, K$. The spectrum then is synonymous with the eigenvalues $\{ \lambda_l \}$. The corresponding right eigenvectors $\{ \psi_l \}$ can be used to embed the system in a low-dimensional representation (or CVs).

Based on this eigendecomposition, the target mapping $\xi(\bx)$ (\req{eq:collective-variables}) can be defined as {\it diffusion coordinates}~\cite{coifman2005geometric,coifman2006diffusion}:
\begin{equation}
  \label{eq:diffusion-coordinates}
  \bx \mapsto \xi(\bx) = \ppar[\big]{\lambda_1 \psi_1(\bx), \dots, \lambda_d \psi_d(\bx)},
\end{equation}
where the eigenvalues and eigenvectors are given by $\pset{\lambda_l}$ and $\pset{\psi_l}$, respectively, and define reduced coordinates. In \req{eq:diffusion-coordinates}, each diffusion coordinate is defined as $z_k = \lambda_k \psi_k$, where the spectrum is sorted by non-increasing value:
\begin{equation}
  \label{eq:eigenvalues}
  \lambda_0=1 > \lambda_1 \ge \dots \ge \lambda_d \ge \dots \ge \lambda_K,
\end{equation}
where $d$ is the index at which we truncate the diffusion coordinates in \req{eq:diffusion-coordinates} and the dimensionality of the reduced representation. Thus, it is expected that the dominant timescales found in the dynamics of the high-dimensional system can be described only by several eigenvectors corresponding to the largest eigenvalues. As the dynamics in DMAP is represented by the transition probability matrix (\req{eq:diffusion-map-markov}), the eigenvalue $\lambda_0=1$ and the first diffusion coordinate $\lambda_0\psi_0$ corresponds to the Boltzmann equilibrium distribution. Therefore, we exclude it from the target mapping (\req{eq:diffusion-coordinates}).

The truncation in \req{eq:eigenvalues} can be justified as follows. As atomistic systems are usually metastable (\rsct{sec:enhanced-sampling}), it is sufficient to approximate the diffusion coordinates by a set of dominant eigenvalues (e.g., up to $d$ in \req{eq:diffusion-coordinates} and \req{eq:eigenvalues}). This corresponds to a negligible error on the order of $O(\lambda_{d}/\lambda_{d-1})$. In other words, a sufficient condition for this is $\lambda_{d-1} \gg \lambda_{d}$ as it relates to a large {\it spectral gap}. As such, the spectral gap separates slow degrees of freedom (for $\ge \lambda_d$) and fast degrees of freedom (for $<\lambda_{d}$).

\subsubsection{Diffusion and Commute Distances}
In the reduced space of the diffusion coordinates, we can define the diffusion distance, a measure of proximity for the samples lying on a low-dimensional manifold. The diffusion distance is equivalent to Euclidean distance on the manifold~\cite{coifman2005geometric}:
\begin{align}
  \label{eq:diffusion-distance}
  D^2(\bx_k,\bx_l) =
    \sum_{n=1}^d \lambda_n^2 \ppar[\big]{\psi_n(\bx_k) - \psi_n(\bx_l)}^2 =
    \|\xi(\bx_k) - \xi(\bx_l)\|^2,
\end{align}
where the definition of $\xi(\bx)$ is given by \req{eq:diffusion-coordinates}. Alternatively, we can write that the diffusion distance between high-dimensional samples $\bx_k$ and $\bx_l$ is equivalent to Euclidean distance between CV samples $\bz_k = \xi(\bx_k)$ and $\bz_l = \xi(\bx_l)$.

The diffusion coordinates (\req{eq:diffusion-coordinates}) can also be defined using the relation of the effective timescales and the eigenvalues~\cite{nadler2006diffusion}. Employing the fact that the eigenvalues decay exponentially:
\begin{equation}
  \label{eq:effective-timescale}
  \lambda_n(\tau)=\e^{-\tau \kappa_n}=\e^{-\tau/ t_n},
\end{equation}
where $\tau$ is a lag time, $\kappa_n$ denotes relaxation rates and $\kappa_n^{-1}=t_n$ are effective timescales, we can rewrite \req{eq:diffusion-coordinates} to obtain kinetic map:
\begin{equation}
  \label{eq:lagged-target-mapping}
  \xi(\bx) = \ppar[\big]{\e^{-\tau/t_1} \psi_1(\bx), \dots, \e^{-\tau/t_d} \psi_d(\bx)},
\end{equation}
where $\tau$ should be selected so the samples are uncorrelated. Selecting the lag time is a known issue in modeling metastable dynamics. Namely, the target mapping in \req{eq:lagged-target-mapping} strongly depends on the lag time $\tau$. The lag time should be selected to separate fast and slow processes if there is an evident timescale separation, i.e., the lag time value is selected between fast and slow timescales. However, these timescales are often unknown before running simulations.

Relaxation timescales (\req{eq:effective-timescale}) can be calculated only when a time-lagged construction of pairwise transition probabilities is used. For instance, this can be done using Mahalanobis distance as a distance metric which employs estimating a time covariance matrix~\cite{dsilva2015data}, the Taken theorem (a delay embedding theorem)~\cite{packard1980geometry}, or the von Neumann entropy~\cite{moon2019visualizing}.

A particularly interesting method that does not require selecting the lag time relies on integrating \req{eq:diffusion-distance} over the lag time $\tau$. Then, we use the relation between the eigenvalues and the effective timescales (\req{eq:effective-timescale}) to arrive at commute distances~\cite{noe2016commute}:
\begin{align}
  \label{eq:commute-distances}
  D_c^2(\bx_k,\bx_l)
    = \sum_{n=1} \pbkt*{\ppar[\big]{\psi_n(\bx_k) - \psi_n(\bx_l)}^2 \int_0^{\infty}\dd{\tau}\e^{-\tau/t_n}}
    = \frac{1}{2}\sum_{n=1}t_n\ppar[\big]{\psi_n(\bx_k) - \psi_n(\bx_l)}^2.
\end{align}
\req{eq:commute-distances} is approximately the average time the systems spends to commute between $\bx_k$ and $\bx_l$, and the associated commute map is given by:
\begin{equation}
  \xi(\bx) = \ppar*{ \sqrt{\frac{t_1}{2}} \psi_0(\bx), \dots, \sqrt{\frac{t_d}{2}}\psi_d(\bx)},
\end{equation}
where we can see the difference between the diffusion and commute distances are in the coefficients $\lambda_n$ and $t_n/2$, respectively. Various methods exploit the relation of the effective timescale with eigenvalues~\cite{noe2015kinetic,boninsegna2015investigating,noe2016commute,klus2018data,banisch2020diffusion,tsai2021sgoop,evans2022computing,evans2023computing,rydzewski2023selecting}.

\chg{
\subsubsection{Diffusion Reweighting}
\label{sec:diffusion-reweighting}
\begin{figure}
  \includegraphics{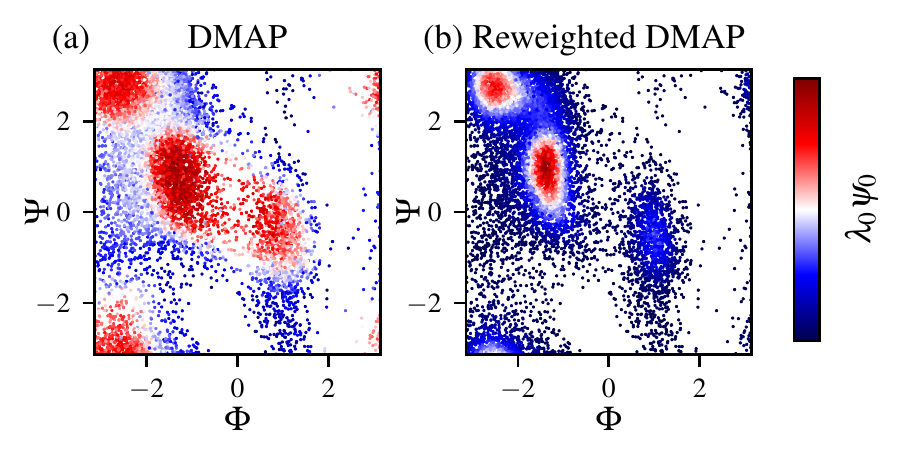}
  \caption{{\bf Diffusion Reweighting}. The equilibrium diffusion coordinate $\lambda_0\psi_0$ calculated for an alanine dipeptide dataset at a temperature of 300 K in vacuum generated using well-tempered metadynamics with a bias factor of 5. The dataset consists of 5000 samples, with 45 features being all pairwise heavy-atom distances in the system. The standard DMAP captures a biased low-dimensional manifold that does not correspond to the equilibrium distribution, as seen by the lowered free-energy barriers and consequently boosted transitions between the metastable states. In contrast, the reweighted DMAP correctly reverts the effect of sampling from a biased probability distribution.}
  \label{fig:diffusion-reweighting}
\end{figure}

The standard DMAP can construct CVs from unbiased atomistic simulations (\rsct{sec:manifold-learning}). However, an approach incorporating statistical sample weights is necessary to learn from enhanced sampling simulations where sampling follows a biased probability distribution. Several unbiasing algorithms are available for DMAP, such as target-measure DMAP~\cite{banisch2020diffusion,trstanova2020local} and its variant that uses Mahalanobis distances~\cite{evans2023computing}. 

Here, we focus on recently proposed reweighted DMAP, which implements \emph{diffusion reweighting} to unbias Markov transitions~\cite{rydzewski2022reweighted,rydzewski2023selecting}. Reweighted DMAP employs a weighted Markov transition matrix:
\begin{equation}
    M(\bx_k, \bx_l) = \frac{r(\bx_k, \bx_l) G_\varepsilon(\bx_k, \bx_l)}{\sum_n r(\bx_k, \bx_n) G_\varepsilon(\bx_k, \bx_n)},
\end{equation}
that uses a pairwise reweighting factor $r(\bx_k, \bx_l)$ to weight each Gaussian. It can be shown that the pairwise reweighting factor takes a simple form~\cite{rydzewski2022reweighted,rydzewski2023selecting}:
\begin{align}
  r(\bx_k,\bx_l) = \frac{w(\bx_k)}{{[\varrho(\bx_k)]^\alpha}} \frac{w(\bx_l)}{{[\varrho(\bx_l)]^\alpha}},
\end{align}
where $\varrho(\bx_k)=\sum_l w(\bx_l) G_\varepsilon(\bx_k,\bx_l)$ is up to a multiplicative constant an unbiased density estimator at sample $\bx_k$. For a detailed derivation and variants of the reweighting factor used in other manifold learning techniques, we refer to~\rref{rydzewski2022reweighted}.

A procedure for constructing reweighted DMAP can be implemented according to Algorithm~\ref{alg:diffusion-map}. We show how this technique can be employed to obtain an equilibrium density from a biased simulation and a comparison to the standard DMAP in \rfig{fig:diffusion-reweighting}.
\begin{algorithm}
  \label{alg:diffusion-map}
  \SetKwInOut{Input}{Input}
  \SetKwInOut{Output}{Output}
  \Input{Dataset $\pset*{\bx_k}_{k=1}^K$, anisotropic diffusion constant $\alpha$.}
  \Output{Eigenvalues $\pset{\lambda_k}$ and eigenvectors $\pset{\psi_k}$ of the transition matrix $M$.}
  \begin{enumerate}[leftmargin=0cm]
    \item Calculate the squared pairwise distances $\| \bx_l - \bx_l \|^2$.
    \item Estimate the Markov transition matrix $M(\bx_k,\bx_l)$:
    \begin{enumerate}[leftmargin=0.6cm]
      \item Calculate the Gaussian kernel $G_\varepsilon(\bx_k,\bx_l)$ (\req{eq:gaussian-kernel}).
      \item Construct the anisotropic diffusion kernel including the anisotropic diffusion constant $\alpha$ (\req{eq:diffusion-map-kernel}) and diffusion reweighting if the dataset is sampled from a biased probability distribution (\req{eq:reweighting-factor}).
      \item Normalize to obtain the row-stochastic Markov transition matrix $M(\bx_k,\bx_l)$ (\req{eq:diffusion-map-markov}).
    \end{enumerate}
    \item Perform eigendecomposition $M\psi_k=\lambda_k\psi_k$ and based on the spectral gap and dominant eigenvalues estimate the diffusion coordinates $\xi(\bx)$ (\req{eq:diffusion-coordinates}).
  \end{enumerate}
  \caption{Reweighted Diffusion Map}
\end{algorithm}
}

\subsubsection{Applications}

\chg{
DMAP and its variants have been used extensively to study the properties of complex systems. Noteworthy applications include constructing CVs and low-dimensional representations from unbiased simulations~\cite{nadler2006diffusion,coifman2008diffusion,singer2009detecting,ferguson2010systematic,ferguson2011nonlinear,rohrdanz2011determination,rohrdanz2013discovering,kim2015systematic,long2019landmark}, estimating free-energy landscapes~\cite{zheng2013rapid,zheng2013molecular,chiavazzo2017intrinsic}, modeling kinetics~\cite{noe2015kinetic,boninsegna2015investigating,noe2016commute}, constructing reweighted manifolds and CVs from biased simulations~\cite{ferguson2011integrating,zhang2018unfolding,banisch2020diffusion,trstanova2020local,rydzewski2022reweighted}, and selecting high-dimensional representations~\cite{rydzewski2023selecting}. DMAP is implemented in the \texttt{pydiffmap} package~\cite{pydiffmap}.
  }

\subsection{Time-Lagged Independent Component Analysis (TICA)}
\label{sec:tica}
TICA was introduced by Molgedey and Schuster in 1994 to determine the mixing coefficients of linearly superimposed uncorrelated signals~\cite{molgedey1994separation}. This method has inspired the development of novel techniques to analyze long simulation trajectories by considering them as propagating signals~\cite{alakent2004application,hernandez2013identification,schwantes2013improvements,endo2018multi,tsai2020learning,wu2020variational}. In 2013 two research groups independently applied TICA in Markov state modeling to understand the kinetics of conformational changes~\cite{hernandez2013identification,schwantes2013improvements}. Currently, TICA is widely used to project high-dimensional simulation data to a low-dimensional manifold, which helps to uncover slow molecular processes more effectively~\cite{wu2020variational}.

\subsubsection{Markov Propagator of Stochastic Processes}
As in the case of DMAP, we start from the notion that dynamics of the microscopic coordinates of the system can be explained by the stochastic differential equation given by \req{eq:sde}. The corresponding transition probability $p_\tau(\bx,\by)\dx=\operatorname{Prob}\pset[\big]{\bx_{t+\tau} = \by\,|\,\bx_t=\bx}$ describes the likelihood of the transition from $\bx$ to $\by$ in a time step $\tau$. We can describe the evolution of a time-dependent probability distribution as:
\begin{equation}
  \rho_{t+\tau}(\by) = \int\dx p_\tau(\bx,\by) \rho_t(\bx),
\end{equation}
which alternatively can be given by the Markov time-propagator $\cM_{\tau}$:
\begin{equation}
  \rho_{t+\tau}(\bx) = \cM_{\tau} \rho_t(\bx),
\end{equation}
that maps a probability distribution $\rho_t(\bx)$ to $\rho_{t+\tau}(\bx)$. We want to emphasize that the time-propagator $\cM_{\tau}$ is closely related to the semigroup $\cK_\tau$ and infinitesimal generator $\cL$~\cite{wu2020variational}. The semigroup is given as:
\begin{equation}
  {\cK_{\tau} g}(\bx) = \mathbb{E}\pbkt[\big]{g(\bx_{t+\tau})\,|\,\bx_t=\bx},
\end{equation}
where $g$ is an auxiliary function. Then, the generator $\cL$ is defined as $\cL g=\lim_{t\rightarrow 0^+}(\cK_\tau g-g)/t$. The adjoint operator of $\cL$, denoted as $\cL^\dagger$, determines the time-evolution of $\rho$:
\begin{equation}
  \frac{\partial \rho_t(\bx)}{\partial t}=\cL^\dagger \rho_t(\bx),
  \label{eq:fpe}
\end{equation}
which is the famous Fokker-Planck equation. From \req{eq:fpe}, we can see that the time-propagator has the form: 
\begin{equation}
  \cM_{\tau}=\e^{\tau\cL^\dagger}.
  \label{eq:expu}  
\end{equation}

To understand the kinetic characteristics of the system, it is necessary to analyze the eigenvalues and eigenfunctions of the generators $\cL$ and  $\cL^\dagger$, as well as the propagator $\cM_{\tau}$. The eigenvalues and eigenfunctions of $\cL$ satisfy the equation:
\begin{equation}
    \cL \psi_k = -\lambda_k \psi_k,
    \label{eq:l-eigen}
\end{equation}
where the eigenvalues are sorted by increasing values $0=\lambda_0<\lambda_1 \leq \cdots$. The eigenvalues of $\cL^\dagger$ are the same as those of $\cL$, while the eigenfunctions of $\cL^\dagger$ are expressed as $\phi_k = \pi \psi_k$, where $\pi$ is the equilibrium Boltzmann distribution. We can notice that $\pi$ is also an eigenfunction of $\cL^\dagger$ with $\lambda_0=0$, which means that $\psi_0=1$. This is because $\lambda_0=0$ is non-degenerate and corresponds to a unique equilibrium distribution. With $\lambda_k$ and $\phi_k$, we can formally write the time-dependent probability distribution:
\begin{equation}
    \rho_t(\bx) = \pi + \sum_{k=1} A_k \e^{-\lambda_k t} \phi_k(\bx),
    \label{eq:tdprob}
\end{equation}
where coefficients are determined by the initial conditions $A_k = \int\dx (\bx) \psi_k(\bx)\pi$. \req{eq:tdprob} means that $\rho_t(\bx)$ converges to $\pi$ when $t\rightarrow\infty$. The eigenfunctions $\phi_k$ can be viewed as different ``modes'' with an equilibration rate determined by $\lambda_k$, where small $\lambda_k$ suggests a slow equilibration along $\phi_k$. As $\cM_{\tau}$ shares the eigenfunctions with $\cL^\dagger$, while the eigenvalues of $\cM$ are $\lambda^\cM_k = \e^{\tau\lambda_k}$, the equilibration timescale is $t_k=-\tau/\log\lambda_k^\cM$.

\subsubsection{Variational Optimization of Dominant Eigenvectors}
As previously discussed, the eigenfunctions of $\cM_{\tau}$ are useful in defining the target mapping for slow equilibration rates when $\lambda^\cM_k$ are small. The first non-trivial eigenfunction $\psi_1$ is the most significant mode, which corresponds to the smallest non-zero eigenvalue. It can be estimated variationally by maximizing:
\begin{align}
    \pang[\big]{g\,|\,\cM_{\tau}\,|\,g}_{\pi}
    = \int\dx\dy g(\bx) p_\tau(\bx,\by) g(\by) \pi,
    \label{eq:tica-loss}
\end{align}
with a constraint given as $\int\dx g(\bx) \pi=0$, where $g$ is a function to be optimized. \req{eq:tica-loss} can be generalized to obtain the eigenvalues and eigenfunctions of $\cM_{\tau}$~\cite{hernandez2013identification,nuske2014variational}. As $\psi_0=1$, $g(\bx)$ can be expressed as a linear combination:
\begin{equation}
    g(\bx) = \sum_k c_k \eta_k(\bx),
    \label{eq:linear-g}
\end{equation}
where $c_k$ are coefficients and $\eta_k(\bx)$ are basis functions. It can be shown that the maximization of \req{eq:tica-loss} is equivalent to solving a generalized eigenproblem:
\begin{equation}
    M\chi=\lambda^M S\chi,
    \label{eq:geigen}
\end{equation}
where $M_{kl}=\pang[\big]{\eta_i\,|\,\cM_{\tau}\,|\,\eta_j}$, the overlap matrix $S$ is defined as $S_{kl}=\pang[\big]{\eta_k\,|\,\eta_l}$, and $\chi_0$ corresponds to the equilibrium distribution. Features with zero-mean are usually taken as
basis functions~\cite{molgedey1994separation,naritomi2011slow,schwantes2013improvements,hernandez2013identification}. However, other basis functions can also be employed. For instance, indicator functions of a partition of the configuration space~\cite{hernandez2013identification,nuske2014variational} or Gaussian functions can be used, which requires fewer macrostates and shorter $\tau$ to converge~\cite{nuske2014variational}. Neural networks can also be used~\cite{bonati2021deep}.

In practice, $M_{kl}$ depends on the lag time $\tau$, which is often taken to be on a nanosecond timescale~\cite{mcgibbon2015variational,sultan2017tica}. For unbiased simulation data, we can estimate $M_{kl}(\tau)$ as:
\begin{equation}
    M_{kl}(\tau) = \frac{1}{K-K_\tau}\sum_{n=1}^{K-K_\tau} x_k(n) x_l(n+K_\tau),
    \label{eq:c-tau}
\end{equation}
where $x_k$ and $x_l$ are $k-$th and $l$-th features, respectively, and the time lag is given as $\tau=K_\tau\Delta t$, where $\Delta t$ is a time step. As $x_k$ and $x_l$ are usually not orthonormal, the overlap matrix can be estimated as:
\begin{equation}
  S_{kl} = M_{kl}(\tau=0) = \frac{1}{K}\sum_{n=1}^{K}x_k(n) x_l(n),
  \label{eq:c0}
\end{equation}
which reduces the generalized eigenproblem (\req{eq:geigen}) to $M(\tau)\chi=\lambda^M M(0)\chi$. We can see that the matrix $M$ has a different interpretation compared to other methods reviewed here, e.g., rather than describing the probabilities between samples, it is given by autocorrelation functions.

A procedure for performing TICA is given in Algorithm~\ref{alg:tica}.
\begin{algorithm}
  \label{alg:tica}
  \SetKwInOut{Input}{Input}
  \SetKwInOut{Output}{Output}
  \Input{Unbiased dataset $\{\bx_k\}_{k=1}^K$, high-dimensional configuration variables (features) $\bx=\pset{x_k}_{k=1}^n$, preselected lag-time $\tau$.}
  \Output{Eigenfunctions $\pset{\psi_k}_{k=1}^d$.}
  \begin{enumerate}[leftmargin=0cm]
    \item Calculate ensemble average $\pang{\bx}$.
    \item Calculate shifted features $\tilde{\bx}=\bx-\pang{\bx}$.
    \item Calculate $M(\tau)$ with \req{eq:c-tau} and $M(0)$ with \req{eq:c0}.
    \item Solve the generalized eigenproblem with the AMUSE algorithm~\cite{tong1990amuse,hernandez2013identification}. This is mainly because $M(0)$ is often ill-conditioned thus $M(0)^{-1}$ is numerically unstable.
    \begin{enumerate}[leftmargin=0.6cm]
      \item Perform principle component analysis to transform $\tilde{\bx}$ to principle
      components $\mathbf{y}$.
      \item Calculate normalized $\tilde{\mathbf{y}}=\Sigma^{-1}\mathbf{y}$, where $\Sigma$ is a diagonal matrix whose $k-$th diagonal element is the standard deviation of $y_k$.
      \item Calculate $\tilde{M}(\tau)$ with respect to $\tilde{\mathbf{y}}$.
      \item Diaganoalze $\tilde{M}(\tau)$. Use the $k$-th eigenvector as coefficients to construct a linear combination of $\tilde{\mathbf{y}}$ to calculate $\psi_k$.
    \end{enumerate}
  \end{enumerate}
  \caption{Time-Independent Component Analysis}
\end{algorithm}

\chg{
\subsubsection{TICA and Enhanced Sampling Simulations}
Generating TICA directly from enhanced sampling simulation is another approach worth considering. However, this remains challenging as many enhanced sampling methods do not preserve unbiased kinetics. CVs generated with TICA and a Markov state model can be used with metadynamics, which has improved the sampling efficiency of complex systems~\cite{sultan2017tica,sultan2018transferable}. 

A successful example of this is constructing CVs with TICA and well-tempered metadynamics~\cite{jay2017variational,bonati2021deep}. It has been shown that not only can configurations of the system be unbiased~\cite{bonomi2009reconstructing,tiwary2015time,giberti2020iterative} but also an effective timescale $\dd{\tau}$ can be obtained by rescaling the simulation time~\cite{jay2017variational}:
\begin{equation}
    \dd{\tau} = \e^{\beta\pbkt*{V(\bz,t)-c(t)}} \dd{t},
    \label{eq:change-time}
\end{equation}
where $V(\bz)$ is a time-dependent bias potential acting in the CV space (\req{eq:bias-potential}) and $c(t)$ is a time-dependent bias drift~\cite{tiwary2015time}. \req{eq:change-time} is asymptotically correct at the long-time limit, and generally, the rescaling does not correspond to the actual unbiased time.
}

\subsubsection{Applications}
\chg{
TICA and it variants have been used to study molecular kinetics~\cite{mardt2018vampnets,wu2020variational,spiriti2022simulation}, free-energy estimation, finding slow CVs~\cite{hernandez2013identification,wu2017variational,zhang2019improving,bonati2021deep}, and enhanced sampling~\cite{sultan2017tica,jay2017variational,sultan2018transferable}, analysis of complex processes~\cite{schwantes2013improvements,paul2017protein,sultan2018towards,ferruz2018dopamine,ahalawat2018mapping,pantsar2018assessment,mondal2018atomic,sidky2019high,sengupta2019automated,brotzakis2019accelerating,tran2019dissociation,abella2020markov,pantsar2020kras,barros2021markov,song2021folding,wang2021effect,jones2021determining,lohr2022small}. For more applications, we refer to reviews~\cite{chodera2014markov,shukla2015markov,husic2018markov}. TICA is implemented in MSMBuilder~\cite{beauchamp2011msmbuilder2} and PyEMMA~\cite{scherer2015pyemma}.
}

\chg{
\subsection{Spectral Gap Optimization of Order Parameters (SGOOP)}
\label{sec:spectral-gap-optimization}
As shown in the case of DMAP, the timescale separation between fast and slow dynamics can be used to parametrize manifolds for metastable systems. However, sometimes it is sufficient to know the difference between the slow and fast eigenvalues rather than the related eigenvectors based on which a low-dimensional manifold is constructed in DMAP. The spectral gap can be used to estimate the degree of the separation between the fast and slow processes. Consequently, the maximal spectrum gap corresponds to the most optimal construction of the slow variables. An approach for this is based on the maximum entropy framework and referred to as spectral gap optimization of order parameters (SGOOP)~\cite{tiwary2016spectral}.

\subsubsection{Maximizing the Spectral Gap} 
In SGOOP, we assume the dynamics of the configuration variables $\bx = \pset{x_k}_{k=1}^n$ (i.e., order parameters~\cite{tiwary2016spectral}) is Markovian. These configurational variables can be linear or nonlinear functions of the microscopic coordinates, and the space they span is high-dimensional. We then consider a target mapping $\xi(\bx)$ that embeds the $n$-dimensional configuration space $\bx$ into a $d$-dimensional CV space $\bz$, creating a trial CV. We consider a single CV where the target mapping is a linear combination of the configurational variables~\cite{tiwary2016spectral}:
\begin{equation}
  \label{eq:sgoop_linearmapping}
  \xi(\bx) = \sum_{k=1}^{n} c_k x_k, 
\end{equation}
where $c_k$ are adjustable parameters. More generally, the trial CV can be multidimensional, and the target mapping given by a nonlinear mapping~\cite{tiwary2016spectral}.

The CV space is then discretized in a grid. Next, the Markov transition matrix between grid bins $\bz_{k}$ and $\bz_{l}$ in the low-dimensional CV space is defined as:
\begin{equation}
 \label{eq:markov-entropy}
 p_{kl} \sim M(\bz_k,\bz_l) = \e^{\tau K(\bz_k,\bz_l)},
\end{equation}
where $K$ is the rate matrix, and $k$ and $l$ are bin indices. The transition probabilities $p_{kl}$ should not depend on the lag time $\tau$ when it is sufficiently small and the transition matrix is Markovian. As the Markov transition matrix is defined in the CV space, not the high-dimensional configuration space, SGOOP differs from the other methods considered in this review. 

In the maximum caliber framework (a generalization of maximum entropy to dynamics), a path entropy can be defined based on the probability of micropaths for a Markovian process discrete in time and space~\cite{dixit2015inferring,ghosh2020maximum}, which for the CV space is defined as:
\begin{equation}
 \label{eq:sgoop-entropy}
 S = -\sum_{kl} \rho(\bz_k) p_{kl} \log p_{kl},
\end{equation}
where $\rho(\bz_k)$ is the stationary probability of the grid bin $\bz_k$. The path ensemble average of a time-dependent quantity $A_{kl}$ is then calculated as:
\begin{equation}
  \label{eq:sgoop_path_ensemble}
 \langle A \rangle = \sum_{kl} \rho(\bz_k) p_{kl} A_{kl}.
\end{equation}

The path entropy given by \req{eq:sgoop-entropy}, along with constraints based on our knowledge of certain static or dynamical quantities $\langle A^{(n)} \rangle$, and other conditions such as detailed balance, are together called caliber~\cite{tiwary2016spectral,dixit2015inferring,ghosh2020maximum}. Then, by maximizing the caliber, we can determine that the Markov transition matrix is related to stationary probabilities:
\begin{equation}
 \label{eq:entropy-markov}
 M(\bz_k,\bz_l) = \sqrt{\frac{\rho(\bz_l)}{\rho(\bz_k)}} \exp\ppar*{-\sum_n l_n A^{(n)}_{kl}},
\end{equation}
where the sum is over the known static or dynamical constraints $A^{(n)}_{kl}$ for the grid bin $\bz_k$ and $\bz_l$, weighted by Lagrange multipliers $l_n$. It has been shown that maximizing the caliber is equivalent to being the least committal about missing information~\cite{dixit2015inferring,tiwary2016spectral}.

Refs.~\cite{tiwary2016spectral,tiwary2017predicting} consider a single dynamical constrain $\langle N \rangle$ that is the average number of transitions to nearest neighbors observed in time $\tau$ (so that $N_{kl}=1$ if grid bins are nearest neighbors and zero otherwise~\cite{tiwary2017predicting}). In this case, the Markov transition matrix becomes:
\begin{equation}
  \label{eq:entropy-markov2}
  M(\bz_k,\bz_l) = \sqrt{\frac{\rho(\bz_l)}{\rho(\bz_k)}} \exp\ppar*{-l} = \frac{\langle N \rangle} {\sum_{ij} \sqrt{\rho(\bz_{i}) \rho(\bz_{j})}} \sqrt{\frac{\rho(\bz_l)}{\rho(\bz_k)}},
 \end{equation}
where the Lagrange multiplier $l$ is calculated by inserting this equation in \req{eq:sgoop_path_ensemble}~\cite{tiwary2017predicting}. Similar equations can be derived by considering other dynamical observables~\cite{tiwary2017predicting}. 

\req{eq:entropy-markov2} is the main equation of SGOOP. Using this equation, we can calculate the Markov transition matrix $M$ from the equilibrium CV distribution $\rho(\bz)$ for the given trial CV (we can even ignore the Lagrange multiplier $l$ as it only sets an overall timescale that is not important~\cite{tiwary2016spectral}). Furthermore, the equilibrium CV distribution $\rho(\bz)$ is obtained through reweighting, so SGOOP is directly applicable to enhanced sampling simulations biasing any CVs (that can be different from the trial CV).

We can then obtain the eigenvalues $\pset{\lambda_k}$ of the Markov transition matrix $M$ for a given trial CV. Then, by varying the trial CV, for instance, by changing the parameters $c_k$ in \req{eq:sgoop_linearmapping}, we can obtain the spectral gap $\lambda_{s}-\lambda_{s+1}$ between the slowest $s$ processes in the system with the largest eigenvalues and other fast processes for the different trial CVs. The optimal CV is the one that maximizes the spectral gap. Note that we do not need to perform additional biased simulations to obtain the spectral gap for the trial CV, as we can employ reweighting to estimate the equilibrium CV distribution $\rho(\bz)$. 

The framework offered by maximum caliber and SGOOP is versatile and can be expanded in multiple ways. For instance, SGOOP can be extended to multidimensional CVs using conditional probability factorization, as shown in \rref{smith2018multi}. Additionally, SGOOP can be combined with commute distances (\req{eq:commute-distances})~\cite{tsai2021sgoop}. Lastly, SGOOP can be employed to assess CVs obtained through other machine-learning techniques~\cite{10.1063/5.0030931}.

\subsubsection{Applications}

SGOOP has been employed in various applications to atomistic simulations, including protein-ligand interactions~\cite{tiwary2016wet,tiwary2017molecular,smith2018multi,pramanik2019can,10.1002/anie.202200983} and nucleation~\cite{tsai2019reaction,zou2021toward}. A code for SGOOP is available at \url{https://github.com/tiwarylab/SGOOP}.
}

\section{Target Mapping (II): Divergence Optimization}
\label{sec:tm-divergence-optimization}
This section discusses manifold learning methods that use divergence optimization to find the parametric target mapping. In the following, we describe parametric variants of well-known methods such as stochastic neighbor embedding (SNE) (\rsct{sec:sne}) and uniform manifold approximation and projection (UMAP) (\rsct{sec:umap}), which enable learning CVs from unbiased atomistic simulations. Additionally, we cover multiscale reweighted stochastic embedding (MRSE) (\rsct{sec:mrse}) and stochastic kinetic embedding (StKE) (\rsct{sec:stke}), which can be used to learn from enhanced sampling simulations.

\subsection{Stochastic Neighbor Embedding (SNE)}
\label{sec:sne}

\chg{  
Introduced by Hinton and Roweis~\cite{hinton2002stochastic}, SNE is used in many fields where dimensionality reduction is required. Despite its widespread popularity, only a few investigations have explored the theoretical background of SNE, unlike DMAP or Laplacian eigenmap. However, a non-parametric SNE has been recently considered from a theoretical standpoint~\cite{shaham2017stochastic,arora2018analysis,linderman2019clustering,yang2021t}, revealing that SNE methods can effectively separate clusters of data in a low-dimensional embedding if there are well-separated clusters in a high-dimensional space~\cite{shaham2017stochastic,linderman2019clustering}. This result, however, has been suggested to be insufficient to show that SNE can accurately find a low-dimensional representation~\cite{arora2018analysis,yang2021t}. A recent study has also shown an interesting connection between SNE and spectral embedding techniques~\cite{carreira2010elastic,linderman2019clustering}. 

To the most known variants of SNE algorithms, we can include $t$-distributed SNE~\cite{maaten2008visualizing}, parametric $t$-SNE~\cite{maaten2009learning,van2014accelerating}, heavy-tailed symmetric SNE~\cite{yang2009heavy}, perplexity-free $t$-SNE, and fast-interpolation SNE~\cite{linderman2019fast}. In addition, SNE inspired many manifold learning methods, which stems from the fact that SNE comprises several steps that can be solved differently. As such methods, we can consider UMAP~\cite{mcinnes2018umap}, SHEAP~\cite{shires2021visualizing}, and more focused on atomistic simulations, stochastic kinetic embedding (StKE)~\cite{zhang2018unfolding,chen2021collective,rydzewski2022reweighted} and multiscale reweighted stochastic embedding (MRSE)~\cite{rydzewski2021multiscale,rydzewski2022reweighted}.  
}

\subsubsection{Parametric Target Mapping}
We consider a more versatile approach called the parametric SNE, which employs a parametric mapping from a high-dimensional space to a low-dimensional manifold~\cite{maaten2009learning}. This variant uses the target mapping consisting of $d$ functions given by (see also \req{eq:div-opt-mapping}):
\begin{equation}
  \label{eq:parametric-target-mapping}
  \bx \mapsto \xi_{\boldsymbol{\theta}}(\bx) = \pset[\big]{ \xi_k(\bx;\btheta) }_{k=1}^d,
\end{equation}
where parameters $\btheta = \{ \theta_k \}$ are adjusted to correspond to a minimum of divergence. The parametric version offers more flexibility in constructing low-dimensional spaces than the standard SNE. The target mapping can be represented by a neural network, where each layer gradually decreases the dimensionality of the configuration space~\cite{hinton2006reducing}.

\subsubsection{Markov Transitions in High-Dimensional Space}
\label{sec:sne-markov}
In SNE, the Markov transition matrix is constructed using a Gaussian kernel $G_\varepsilon$ (\req{eq:gaussian-kernel}). Then, the corresponding transition probabilities $p_{kl}$ are defined by normalizing each row of the matrix created using the Gaussian kernel with an additional selection of scale parameters (bandwidths):
\begin{equation}
  \label{eq:sne-affinity}
  p_{kl} \sim M(\bx_k,\bx_l) = \frac{G_{\varepsilon_k}(\bx_k,\bx_l)}{\sum_n G_{\varepsilon_k}(\bx_k,\bx_n)},
\end{equation}
where the diagonal elements are set to zero in order to define a non-lazy Markov chain. Each scale parameter $\{ \varepsilon_k > 0\}_{k=1}^K$ corresponds to a row of $M$. The scale parameters can be written as $\varepsilon_k \propto \sigma^2_k$, where $\sigma_k$ is the standard deviation of the Gaussian kernel. Finally, the Markov transition matrix is symmetrized~\cite{van2009dimensionality}.

Selecting $\{ \varepsilon_k \}$ is the main difference compared to constructing the Markov transition matrix from the single-scale Gaussian kernel (\req{eq:gaussian-kernel}) as in DMAP. To find $\{ \varepsilon_k \}$, the Shannon entropy $H(\bx_k)$ of each row of the Markov transition matrix should be optimized to hold the following condition:
\begin{equation}
  \label{eq:sne-entropy}
  H(\bx_k) = \log_2 P,
\end{equation}
where $P$ is a measure of neighborhood at $\bx_k$ referred to as perplexity. Therefore, the optimization task is to adjust $\{ \varepsilon_k \}$ so that $H(\bx_k)$ is roughly $\log_2 P$. After the optimization, the Markov transition matrix contains information on the geometric relationships between samples in the high-dimensional space. In most SNE algorithms, the optimization is performed using logarithmic search. 

\chg{
Perplexity is a concept that considers both local and global aspects of data. However, its value can have a complicated impact on the final result, causing confusion and potentially affecting how we interpret the manifold, particularly in complex physical systems with multiple metastable states of varying densities. Therefore, it is common to construct manifolds with different perplexity values and select a manifold that can be interpreted. For more information on choosing the appropriate perplexity and its impact on low-dimensional representations, we refer to \rref{wattenberg2016use}.}

\begin{figure}
  \includegraphics{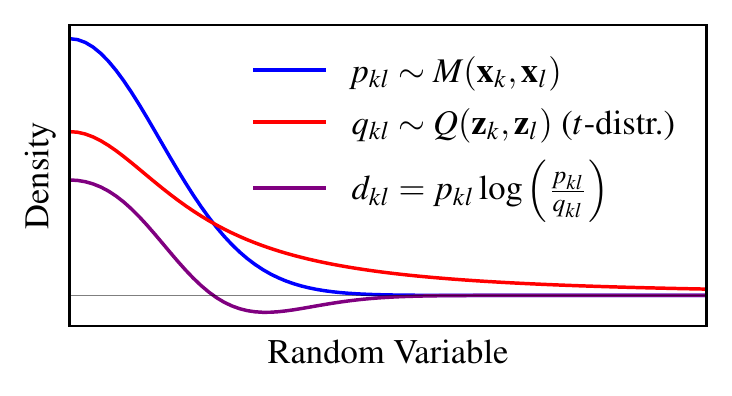}
  \caption{{\bf Crowding Problem.} The approach taken by $t$-distributed SNE to construct Markov transition probabilities from low-dimensional samples. The Markov transition matrix with probabilities $p_{kl}$ between high-dimensional samples is modeled using a Gaussian kernel (blue). In contrast, its equivalent for low-dimensional samples $Q$ is given by a one-dimensional $t$-distribution (red). The difference between $M$ and $Q$ (purple) as measured by each term $d_{kl}=p_{kl}\log\ppar*{p_{kl}/q_{kl}}$ that summed gives the Kullback--Leibler divergence $D_{\mathrm{KL}}=\sum_{kl}d_{kl}$. In complex systems, positive values of $d_{kl}$ increase the propensity to group samples within metastable states, while negative values improve the separation between metastable states.}
  \label{fig:crowding}
\end{figure}

\subsubsection{Manifold Representation}
Depending on the variant of SNE, the Markov transition matrix built from low-dimensional samples can be represented by different normalized kernels $Q(\bz_k,\bz_l)$. In SNE, the transition probabilities $q_{kl}$ between the low-dimensional samples are given by the Gaussian kernel:
\begin{equation}
  \label{eq:q-gaussin-kernel}
  q_{kl} \sim Q(\bz_k,\bz_l) = \frac{\exp\ppar*{ -\frac{1}{\varepsilon}\| \bz_k - \bz_l \|^2 }}{\sum_n \exp\ppar*{ -\frac{1}{\varepsilon}\| \bz_k - \bz_n \|^2 }},
\end{equation}
where the samples are determined by the parametric target mapping such that $\bz=\xi_{\btheta}(\bx)$. A single value of $\varepsilon$ for every row of $Q$ is chosen, which differs from estimating bandwidths to build the Markov transition matrix from high-dimensional samples.

In $t$-distributed SNE~\cite{maaten2008visualizing,maaten2009learning} (abbreviated as $t$-SNE), which is perhaps the most commonly used variant of SNE, $Q(\bz_k,\bz_l)$ is instead represented by a non-parametric $t$-distribution kernel with one degree of freedom (i.e., the Lorentz function) (\rfig{fig:crowding}):
\begin{equation}
  \label{eq:q-t-kernel}
  q_{kl} \sim Q(\bz_k,\bz_l) = \frac{\ppar*{ 1 + \| \bz_k - \bz_l \|^2 }^{-1}}{\sum_n \ppar*{ 1 + \| \bz_k - \bz_n \|^2 }^{-1}},
\end{equation}
which does not require selecting additional parameters. Therefore, in contrast to spectral embedding methods, SNE does not employ eigendecomposition but builds the transition matrix from the low-dimensional samples estimated by the parametric target mapping~\cite{hinton2002stochastic,maaten2008visualizing,maaten2009learning}.

Using the $t$-distribution kernel instead of the Gaussian kernel is motivated by the {\it crowding problem}~\cite{maaten2008visualizing,maaten2009learning} where clusters of samples (or metastable states) are not separated correctly. This is because the low-dimensional space available to accommodate moderately distant samples is not large enough compared to the space available to accommodate nearby samples. It is partly caused by the dimensionality curse~\cite{marimont1979nearest,assent2012clustering}. As the $t$-distribution kernel is heavy-tailed (it is an infinite mixture of Gaussians with different variances), the separation of the clusters is more optimal (\rfig{fig:crowding}). The choice of a one-dimensional $t$-distribution is because $(1+\pnrm{\bz_k-\bz_l}^2)^{-1}$ approaches an inverse square law for large pairwise distances in the low-dimensional representation~\cite{maaten2008visualizing}.

\begin{figure*}
  \includegraphics[width=0.75\textwidth]{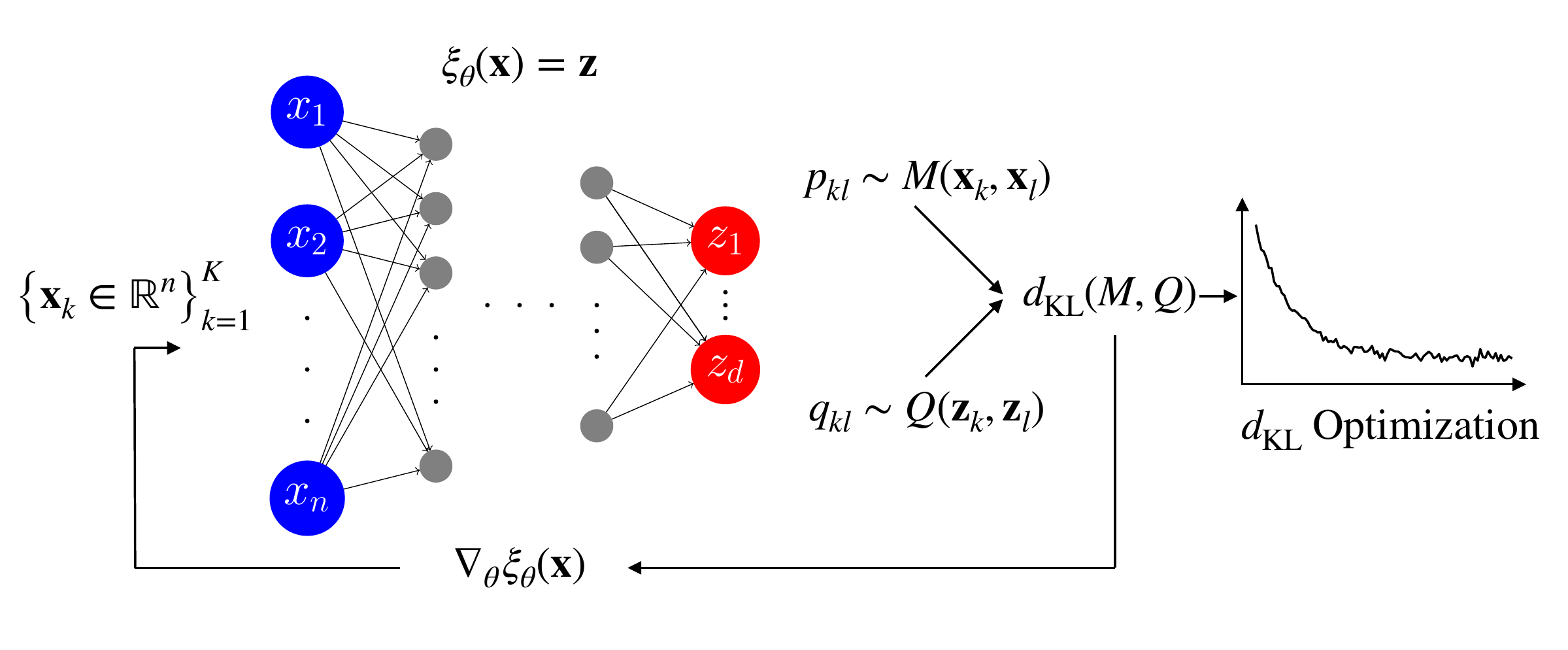}
  \caption{{\bf Parametric Target Mapping}. Schematic depiction of learning parameters $\mathbf{\theta}$ for a parametric target mapping represented by a neural network. The backpropagation procedure estimates errors of the parametric target mapping $\nabla_{\mathbf{\theta}}\xi_{\mathbf{\theta}}(\bx)$ which are used to correct the parameters so that the Kullback--Leibler divergence (or any other divergence) calculated from the Markov transition matrix $M(\bx_k,\bx_l)$ and $Q(\bz_k,\bz_l)$ decreases to zero. A minimum value of the Kullback--Leibler divergence indicates that the relations between samples in the configuration and CV space are preserved.}
  \label{fig:parametric-target-mapping}
\end{figure*}

\subsubsection{Learning Algorithm}
To correctly parametrize the target mapping and CV samples, we aim to achieve a minimal difference between the high-dimensional and low-dimensional Markov transition matrices, $M$ and $Q$, respectively. Many divergences allow computing statistical distances between pairwise probability distributions. SNE uses the Kullback--Leibler divergence~\cite{kullback1951information} reformulated to compare Markov transition matrices row by row~\cite{rached2004kullback}:
\begin{align}
  \label{eq:kl-divergence}
  D_{\mathrm{KL}}\ppar*{M, Q} = \sum_{kl} p_{kl} \log\ppar*{\frac{p_{kl}}{q_{kl}}} =
    \sum_{kl} p_{kl} \log p_{kl} -
    \sum_{kl} p_{kl} \log q_{kl},
\end{align}
where the summation goes over every pair of $k$ and $l$. The first term in \req{eq:kl-divergence} is constant and can be excluded from the minimization. The Kullback--Leibler divergence is greater than 0 and equal to 0 only when $p_{kl} = q_{kl}$ for every pair $(k,l)$. Other divergences can also be used to compare $M$ and $Q$, e.g., the symmetrized Kullback--Leibler or the Jensen--Shannon divergences.

\begin{algorithm}
  \label{alg:parametric-sne}
  \SetKwInOut{Input}{Input}
  \SetKwInOut{Output}{Output}
  \Input{Dataset $\{\bx_k\}_{k=1}^K$, perplexity $P$.}
  \Output{Target mapping $\xi_{\btheta}(\bx)=\pset{\xi_k(\bx; \btheta)}_{k=1}^d$.}
  \begin{enumerate}[leftmargin=0cm]
    \item Sample landmarks if necessary.
    \item Initiate parameters $\btheta$ for the target mapping $\xi_{\btheta}$.
    \item Iterate over training epochs:
      \begin{enumerate}[leftmargin=0.6cm]
        \item Map the samples to their reduced representation: $\xi_{\btheta}: \bx_l \mapsto \bz_l$.
        \item Iterate over data batches:
      \begin{enumerate}[leftmargin=0.6cm]
        \item Calculate the Markov transition matrices $M$ (dependent on perplexity $P$) and $Q$ (dependent on $\btheta$).
        \item Update optimization step with the loss given by the divergence for the updated parameters $\btheta$.
      \end{enumerate}
    \end{enumerate}
  \end{enumerate}
  \caption{Parametric Stochastic Neighbor Embedding}
\end{algorithm}

Algorithm~\ref{alg:parametric-sne} outlines the learning procedure employed in SNE. The neural network representing the parametric target mapping is trained iteratively, which can be performed using many stochastic gradient descent algorithms, e.g., Adam~\cite{kingma2014adam}. Then, backpropagation is used to estimate errors of the parametric target mapping $\nabla_{\mathbf{\theta}}\xi_{\mathbf{\theta}}(\bx)$ which corrects the parameters. This results in the decrease of the Kullback--Leibler divergence (or any other divergence) calculated from the Markov transition matrices $M(\bx_k,\bx_l)$ and $Q(\bz_k,\bz_l)$. Finally, the minimum value of the Kullback--Leibler divergence indicates that the geometric relations between samples in the configuration and CV space are preserved (\rfig{fig:parametric-target-mapping}).

\subsubsection{Applications}
\chg{
For atomistic simulations, SNE and its extensions have been applied to constructing CVs and low-dimensional representations~\cite{rydzewski2016machine,rydzewski2017ligand,zhou2018t,rydzewski2021multiscale}, clustering metastable states~\cite{shires2021visualizing,nicoli2022classification,appadurai2022demultiplexing}, and analyzing molecular structures~\cite{romero2019mechanism}. $t$-SNE is implemented in the \texttt{sklearn} library~\cite{sklearn}.
}

\subsection{Uniform Manifold Approximation and Projection (UMAP)}
\label{sec:umap}

\chg{
UMAP is a method for unsupervised manifold learning first introduced by McInnes et al. in 2018~\cite{mcinnes2018umap}. Recently, a parametric variant of UMAP has been developed~\cite{sainburg2021parametric}. UMAP is now considered on par with $t$-SNE with applications in various disciplines. Similar to Laplacian eigenmap~\cite{belkin2001laplacian,belkin2003laplacian}, UMAP builds on the assumption that a high-dimensional dataset is uniformly distributed on a low-dimensional manifold. However, this assumption has been challenged for many datasets, even those where UMAP performs well, making it difficult to understand the real reason for its success~\cite{damrich2021umap}. 

For an easy comparison to other parametric manifold learning techniques discussed in this review, we cover only the general aspects of UMAP, including the construction of transition matrices from high- and low-dimensional samples and divergence minimization. For readers interested in the mathematical concepts involved in UMAP and a detailed comparison to $t$-SNE, we refer to \rref{mcinnes2018umap,damrich2021umap,damrich2022t}.}

\subsubsection{Unnormalized Transition Matrix}
UMAP uses an \emph{unnormalized} transition matrix to represent relations between samples in high-dimensional space. The lack of normalization of the transition probabilities matrix in UMAP is motivated mainly by reducing the computational cost associated with normalization. The Markov transition probability matrix is defined in UMAP as:
\begin{equation}
  \label{eq:umap-markov}
  p_{kl} \sim M(\bx_k,\bx_l) = \exp\ppar*{-\frac{1}{\varepsilon_k} \ppar{\| \bx_k - \bx_l \|^2 - d_k}},
\end{equation}
where the distance from each $k$-th sample to its first neighbor is denoted as $d_k$, ensuring that every probability $p_{kl} \le 1$ and divergence can be used as a loss function. Similar to SNE, the diagonal elements of $M$ are zeroed. Although the Markov transition matrix used by UMAP is very similar to that of SNE (\req{eq:gaussian-markov-matrix}), UMAP uses a different symmetrization scheme~\cite{mcinnes2018umap} instead of symmetrizing the Markov transition matrix as in SNE.

UMAP does not employ perplexity to find the scale parameters (\req{eq:umap-markov}). Instead, based on the selected number nearest neighbors $m$, UMAP estimates the scale parameters $\pset{\varepsilon_k}$ so that $\log_2 m = \sum_l M(\bx_k,\bx_l)$. This can be employed, unlike in SNE techniques, as the Markov transition matrix in UMAP is not row-normalized and $\sum_l M(\bx_k,\bx_l) \neq 1$. Consequently, this optimization procedure does not require the calculation of the Shannon entropy of each row of the Markov transition matrix (\rsct{sec:sne-markov}) as is done in SNE, making UMAP less computationally demanding.

\subsubsection{Transition Probabilities in Manifold}
\begin{figure}
  \includegraphics{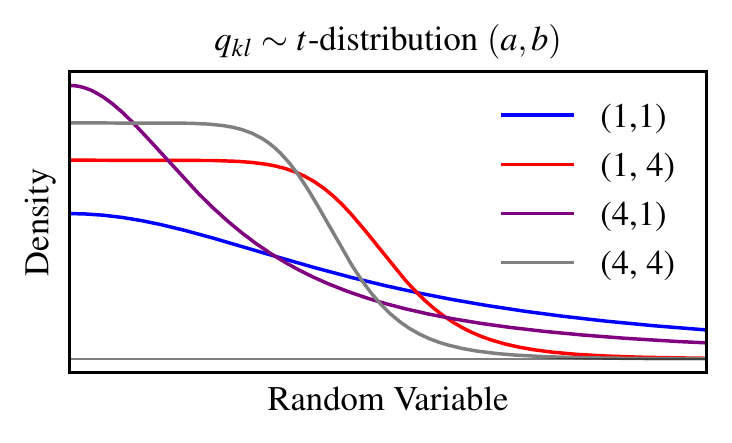}
  \caption{{\bf Low-Dimensional Transition Probabilities $q_{kl}$ in UMAP.} Procedure using a parametrizable $t$-distribution to represent data states in a low-dimensional manifold. This distribution with $(a,b)=(1,1)$ corresponds to the $t$-distribution used in $t$-SNE. UMAP finds $a$ and $b$ using optimization to match the distribution of pairwise distances calculated from the low-dimensional samples.}
  \label{fig:q-umap}
\end{figure}

As in every manifold learning technique relying on divergence optimization, UMAP needs to construct the probability transition matrix from low-dimensional samples (\rfig{fig:q-umap}). To provide more flexibility in separating states in the manifold, the probability transition matrix in UMAP is estimated by the following $t$-distribution kernel:
\begin{equation}
  \label{eq:umap-q}
  q_{kl} \sim Q(\bz_k,\bz_l) \propto \ppar*{1+a\|\bz_k-\bz_l\|^{2b}}^{-1},
\end{equation}
where we omit the normalization factor for brevity, and $a$ and $b$ are parameters that can be determined through an additional optimization procedure. If both parameters are set to one, \req{eq:umap-q} simplifies to the kernel for the low-dimensional probabilities used in $t$-SNE (\req{eq:q-t-kernel}). \chg{In practice, UMAP uses nonlinear least-square optimization to find $a$ and $b$ by fitting \req{eq:umap-q} to a piecewise function:
\begin{equation}
  f_{kl} =
    \begin{cases}
      1 & \text{if $\|\bz_k-\bz_l\|$} < d_m\\
      \exp\ppar{-\|\bz_k-\bz_l\| - d_m} & \text{otherwise},
    \end{cases}       
\end{equation}
where $d_m$ is a constant setting the separation between close samples in a manifold. For instance, the fitting can be performed using the BFGS algorithm~\cite{fletcher2013practical}. In practice, $a$ and $b$ are often constant or fixed after initial learning steps~\cite{mcinnes2018umap}.
}

\subsubsection{Divergence Minimization}
Divergence optimization in UMAP can be performed using a general framework for learning parametric manifold learning techniques (Algorithm~\ref{alg:parametric-sne}). To find a low-dimensional transition matrix and the corresponding target mapping into a manifold, UMAP minimizes a divergence that can be viewed as the sum of cross entropy losses for each transition probability~\cite{mcinnes2018umap}:
\begin{align}
  \label{eq:umap-loss}
  D_{\mathrm{CE}} &= D_{\mathrm{KL}}(M, Q) + D_{\mathrm{KL}}(1-M,1-Q) \nonumber\\
    & = \sum_{kl} p_{kl} \log\ppar*{\frac{p_{kl}}{q_{kl}}}
    + \sum_{kl} (1-p_{kl}) \log\ppar*{\frac{1-p_{kl}}{1-q_{kl}}}
\end{align}
consisting of two Kullback--Leibler divergences $D_{\mathrm{KL}}$ (compare to \req{eq:kl-divergence}) responsible for grouping samples with high transition probabilities in metastable states and separating samples with low transition probabilities into different metastable states, respectively. As in SNE, the loss function can be reduced by excluding constant terms that depend on $p_{kl} \sim M(\bx_k,\bx_l)$. 

\subsubsection{Applications}

\chg{
UMAP has been used to cluster and visualize molecular structures~\cite{shires2021visualizing,trozzi2021umap,oide2022protein,roncoroni2023unsupervised}, and constructing CVs~\cite{le2022behavior}. For examples of applications on simple datasets, we refer to Ref.~\footnote{A.~Coenen, A.~Pearce, Google PAIR \url{https://pair-code.github.io/understanding-umap/}}. Its implementation is available at \url{https://github.com/lmcinnes/umap}.
}

\subsection{Reweighted Stochastic Embedding}
\label{sec:rse}
Reweighted stochastic embedding is a recently developed framework designed to construct low-dimensional manifolds of CVs from standard and enhanced sampling atomistic simulations~\cite{rydzewski2022reweighted}. The framework expands on concepts from DMAP and parametric versions of SNE, with an additional reweighting procedure to account for constructing the Markov transition matrix based on samples yielded from biased probability distributions. To such methods, we can include stochastic kinetic embedding (StKE)~\cite{zhang2018unfolding,chen2021collective} and multiscale reweighted stochastic embedding (MRSE)~\cite{rydzewski2021multiscale,rydzewski2022reweighted}.

\subsubsection{Stochastic Kinetic Embedding (StKE)}
\label{sec:stke}
Stochastic kinetic embedding (StKE) aims to learn the target mapping by matching kinetics encoded in the high-dimensional and low-dimensional spaces~\cite{zhang2018unfolding}. The primary assumption in constructing the unbiased Markov transition matrix from high-dimensional samples is that these samples are uniformly distributed in the configuration space. To achieve this, a sparse set of high-dimensional samples is selected using landmark sampling (\rsct{sec:landmark-sampling}) by requiring that the distance between each pair of landmarks is larger than a predefined threshold value, $\pnrm*{\bx_k - \bx_l} > d_0$. Using the anisotropic diffusion kernel $L$ (\req{eq:diffusion-map-kernel}), the unbiased Markov transition matrix is:
\begin{equation}
    M(\bx_k,\bx_l) = \frac{w(\bx_l) L(\bx_k,\bx_l)}{\sum_n w(\bx_n) L(\bx_k,\bx_n)},
    \label{eq:transition-stke}
\end{equation}
where the statistical weights can be approximated by the unbiased kernel density estimates $w(\bx_k) \approx \varrho(\bx_k)$~\cite{zhang2018unfolding,rydzewski2022reweighted} as the distribution of landmarks virtually uniform. An interesting property that establishes a link between StKE and maximum entropy methods is that \req{eq:transition-stke} can also be written as~\cite{rydzewski2022reweighted}:
\begin{equation}
  \label{eq:sqrt}
  M(\bx_k,\bx_l) = \sqrt{\frac{\varrho(\bx_l)}{\varrho(\bx_k)}} G_\varepsilon(\bx_k,\bx_l),
\end{equation}
where $G_\varepsilon$ is the Gaussian kernel and we take the following approximation $\varrho(\bx_k) \approx \sum_n \varrho(\bx_n) L(\bx_k,\bx_n)$. For a detailed discussion on this approximation, see \rref{rydzewski2022reweighted}.

Apart from constructing the unbiased Markov transition matrix, the remaining elements in StKE are similar to other manifold learning methods that use parametric target mappings. Namely, the low-dimensional transition matrix $Q(\bz_k,\bz_l)$ is constructed using a Gaussian kernel such as in the standard version of SNE (\req{eq:q-gaussin-kernel}). The parametric target mapping $\xi_{\btheta}(\bx)$ is represented as a neural network, and the algorithm learns through minimizing the Kullback--Leibler divergence (\req{eq:kl-divergence}). StKE can be implemented according to Algorithm~\ref{alg:stke}, where we additionally provide a step that divides the learning set into data batches, as is often done for training neural networks.

\begin{algorithm}
  \label{alg:stke}
  \SetKwInOut{Input}{Input}
  \SetKwInOut{Output}{Output}
  \Input{Biased dataset $\{\bx_k\}_{k=1}^K$, preselected $d_0$, scale parameter of the Gaussian kernel $\varepsilon$.}
  \Output{Target mapping $\xi_{\btheta}(\bx)=\pset{\xi_k(\bx; \btheta)}_{k=1}^d$.}
  \begin{enumerate}[leftmargin=0cm]
    \item Reduce the size of the learning set by sampling landmarks:
    \begin{enumerate}[leftmargin=0.6cm]
      \item Randomly select a sample $\bx_k$ and add it to the landmark set.
      \item Iterate over all samples. If a sample $\bx_l$ satisfies $\|\bx_k-\bx_l\|\geq d_0$ for any $\bx_k$ in the landmark set, add $\bx_l$ to the landmark set.
    \end{enumerate}
    \item Calculate $\varrho(\bx_l)$ with kernel density estimation for $\bx_l$ in the landmark set.
    \item Iterate over training epochs:
      \begin{enumerate}[leftmargin=0.6cm]
        \item Map the samples in the landmark set to their reduced representation: $\xi_{\btheta}: \bx_l \mapsto \bz_l$.
        \item Iterate over batches of landmarks from the landmark set:
      \begin{enumerate}[leftmargin=0.6cm]
        \item Calculate the Markov transition matrices $M$ (\req{eq:transition-stke}) and $Q$ (\req{eq:q-gaussin-kernel}).
        \item Update optimization step with the loss function given by the Kullback--Leibler divergence to get the parameters $\btheta$.
      \end{enumerate}
    \end{enumerate}
  \end{enumerate}
  \caption{Stochastic Kinetic Embedding}
\end{algorithm}

\subsubsection{Multiscale Reweighted Stochastic Embedding (MRSE)}
\label{sec:mrse}
As Hinton and Roweis note in their work introducing SNE~\cite{hinton2002stochastic}, the Gaussian representation of the Markov transition matrix used in SNE (\rsct{sec:sne-markov}) can be extended to kernel mixtures, where instead of using a Gaussian kernel with perplexity for determining the scale parameters for each row of the Markov transition matrix, a set of perplexities can be used.

MRSE employs this approach to construct a multiscale Markov transition matrix. In contrast to SNE, we have a range of perplexities $\pset{P_i}$ and, therefore, a matrix of scale parameters $\pset{\boldsymbol{\varepsilon}_k} = \pset{\varepsilon_{ki}}$, where $k$ numbers the row of the Markov transition matrix and $i$ denotes perplexity index. First, we build a Gaussian \emph{mixture} as a sum over Gaussians with different values of the scale parameters:
\begin{equation}
  \label{eq:gaussian-mixture-kernel}
  G_{\boldsymbol{\varepsilon}_{k}}(\bx_k,\bx_l)=\sum_{i} G_{\varepsilon_{ki}}(\bx_k, \bx_l),
\end{equation}
where each $\varepsilon_{ki}$ associated with $\bx_k$ is estimated by fitting to the data, so the Shannon entropy of \req{eq:gaussian-kernel} is approximately $\log_2 P_i$, where $P_i$ is a perplexity value from an automatically selected range. Then, the Gaussian mixture is constructed so that each perplexity corresponds to a different spatial scale, allowing metastable states with different geometry to be characterized. In the next stage, we express the Markov transition matrix by normalizing the Gaussian mixture (\req{eq:gaussian-mixture-kernel}) over many perplexities:
\begin{equation}
  \label{eq:gaussian-mixture-markov}
  p_{kl} \sim M(\bx_k,\bx_l) = \frac{G_{\boldsymbol{\varepsilon}_k}(\bx_k,\bx_l)}{\sum_n G_{\boldsymbol{\varepsilon}_k}(\bx_k,\bx_n)},
\end{equation}
where each entry in the Markov transition matrix is the transition probability from $\bx_k$ to $\bx_l$.

\begin{figure}
  \includegraphics{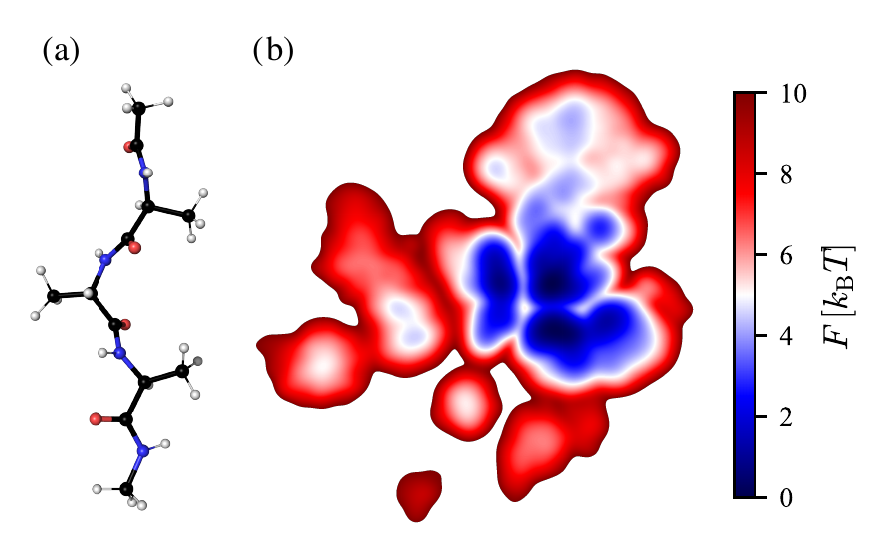}
  \caption{{\bf Free-Energy Landscape of Alanine Tetrapeptide in Vacuum}. Manifold is spanned by CVs calculated using MRSE. The sines and cosines of the $\Psi=(\Psi_1,\Psi_2,\Psi_3)$ and $\Phi=(\Phi_1,\Phi_2,\Phi_3)$ backbone dihedral angles of alanine tetrapeptide are taken as high-dimensional configuration variables. Biased simulation data is generated using well-tempered metadynamics at a temperature of 300 K using $\Psi$ as variables whose fluctuations are enhanced. The WTRS algorithm is used to sample landmarks. For more details, see~\rref{rydzewski2021multiscale}.}
\end{figure}

To learn CVs from biased data from enhanced sampling simulations, we include information about the importance of high-dimensional samples. This this aim, \req{eq:gaussian-mixture-kernel} is redefined as:
\begin{equation}
  \label{eq:reweighting-kernel-mrse}
  G_{\boldsymbol{\varepsilon}_k}(\bx_k,\bx_l) = r(\bx_k,\bx_l) \sum_{i} G_{{\varepsilon}_{ki}}(\bx_k, \bx_l),
\end{equation}
where the pairwise reweighting factor $r(\bx_k,\bx_l)$ needed to unbias the transition probabilities is~\cite{rydzewski2021multiscale}:
\begin{equation}
  \label{eq:reweighting-factor-mrse}
  r(\bx_k,\bx_l)=\sqrt{w(\bx_k) w(\bx_l)},
\end{equation}
which can be shown to be a special case of diffusion reweighting used in reweighted DMAP~\cite{rydzewski2022reweighted}. Using the reweighting factor as a geometric mean of two statistical weights can be justified due to the additive nature of the bias potential.

The remaining components of MRSE are similar to those of StKE. A non-parametric $ t$-distribution represents the low-dimensional transition matrix, and the target mapping is expressed as a neural network that adjusts the parameters of the target mapping so that the Kullback--Leibler divergence is minimal.

\subsubsection{Biasing Manifolds}
A valuable advantage of the parametric target mapping is that high-dimensional samples outside the training set can be mapped to the CV samples. Furthermore, the gradient of the target mapping with respect to the microscopic coordinates $\bx$ can be estimated through backpropagation. This enables using the parametric target mapping as CVs for biasing in enhanced sampling simulations. The additional biasing force acting on the microscopic coordinates at time $t$ is:
\begin{equation}
  \label{eq:biasing-force}
  F(t) = - {\frac{\partial\xi_{\btheta}(\bx)}{\partial\bx}\frac{\partial V(\bz, t)}{\partial\bz}},
\end{equation}
where the second partial derivative is calculated from the bias potential $V$ acting in the CV space defined by the target mapping. The target mapping can also be recomputed during the simulation, allowing iterative improvement of the estimated CVs, which is particularly helpful in insufficiently sampled systems.

\subsubsection{Applications}
MRSE and StKE are recent methods and have been used to construct CVs from standard atomistic and enhanced sampling simulations~\cite{zhang2018unfolding,rydzewski2021multiscale,rydzewski2022reweighted}. MRSE is implemented in an additional module called \texttt{lowlearner}~\cite{rydzewski2021multiscale} in a development version of the open-source PLUMED library~\cite{plumed,plumed-nest} (DOI: \url{https://doi.org/10.5281/zenodo.4756093}).

\section{Conclusions}
\label{sec:conclusions}

\chg{
The use of simulations is widespread for analyzing the dynamics of complex systems at the atomistic level of detail inaccessible for experiments. As a result, we often need to extract meaningful information from observations comprising thousands of configuration variables. Gaining insight into physical processes can be challenging. This problem, however, can be approached using unsupervised manifold learning techniques that can reduce the high-dimensional representations to simplify the data, uncover hidden structures, and describe the dynamics by a few CVs.

Our review provides a unified framework for manifold learning techniques based on building Markov transition matrices from high-dimensional simulation data. Readers more interested in concepts behind a broader class of dimensionality reduction methods can refer to various introductions to the subject, including those from a machine learning perspective~\cite{borg2005modern,lee2007nonlinear,van2009dimensionality,abdi2010principal,ma2012manifold,izenman2012introduction,xie2020representation} or specifically related to atomistic simulations~\cite{noe2017collective,sittel2018perspective,ceriotti2019unsupervised,wang2020machine,bernetti2020data,noe2020machine,gkeka2020machine,glielmo2021unsupervised,chen2021collective,bhatia2023confluence}.

Manifold learning often requires extensive preprocessing of simulation data, including reducing the number of configuration variables to create a high-dimensional space for further dimensionality reduction and sampling landmarks to limit the number of samples in the dataset. These tasks are not trivial and should be performed so that the reduced simulation dataset accurately represents the properties of the system. While many landmark sampling algorithms can be used in standard atomistic and enhanced sampling simulations, selecting the initial high-dimensional representation to be further reduced quantitatively just recently started to gain attention~\cite{ravindra2020amino,rydzewski2023selecting}.

On the other hand, it is also important that have access to well-converged simulation data, which, at this point, is a prerequisite for many manifold learning techniques~\cite{butler2018machine,ceriotti2019unsupervised}. This is critical if extensive sampling is required to reach experimental timescales where many rare processes occur. As such, developing manifold learning techniques often requires testing under idealistic sampling conditions. Several methods circumvent this problem by learning and biasing CVs iteratively at different stages of the progressing enhanced sampling simulation~\cite{chiavazzo2017intrinsic,zhang2018unfolding,ribeiro2018reweighted,chen2018molecular,bonati2020data,bonati2021deep,sun2022multitask,jung2023machine}. This process can benefit from reweighting manifolds constructed from samples generated according to biased probability distributions~\cite{zhang2018unfolding,rydzewski2021multiscale,rydzewski2022reweighted,rydzewski2023selecting}.
}

\section*{Acknowledgements}
J.R. acknowledges funding from the Polish Science Foundation (START), the National Science Center in Poland (Sonata 2021/43/D/ST4/00920, ``Statistical Learning of Slow Collective Variables from Atomistic Simulations''), and the Ministry of Science and Higher Education in Poland. M.C. acknowledges the Purdue Startup Funding. O.V. acknowledges the support of University of North Texas Startup Funding. 

\bibliography{main.bib}

\end{document}